\documentclass[iop,apjl]{emulateapj}

\usepackage{natbib}
\slugcomment{To appear in The Astrophysical Journal}
\shorttitle{The \ubvri\ colors of the Sun}
\shortauthors{Ram\'irez et al.}

\newcommand{\feh}{\mathrm{[Fe/H]}}
\newcommand{\teff}{T_\mathrm{eff}}
\newcommand{\logg}{\log g}

\newcommand{\fei}{Fe\,\textsc{i}}
\newcommand{\feii}{Fe\,\textsc{ii}}

\newcommand{\ubvri}{UBV(RI)$_\mathrm{C}$}

\begin{document}

\title{THE \ubvri\ COLORS OF THE SUN}

\author{I.\,Ram\'irez    \altaffilmark{1},
        R.\,Michel       \altaffilmark{2},
        R.\,Sefako       \altaffilmark{3},
        M.\,Tucci Maia   \altaffilmark{4,5},
        W.\,J.\,Schuster \altaffilmark{2}, \\
        F.\,van Wyk      \altaffilmark{3},
        J.\,Mel\'endez   \altaffilmark{5},
        L.\,Casagrande   \altaffilmark{6}, and
        B.\,V.\,Castilho \altaffilmark{7}
        }
\altaffiltext{1}{McDonald Observatory and Department of Astronomy,
                 University of Texas at Austin, 1 University Station, C1400
                 Austin, Texas 78712-0259, USA}
\altaffiltext{2}{Observatorio Astron\'omico Nacional, 
                 Universidad Nacional Aut\'onoma de M\'exico,
                 Apartado Postal 877, Ensenada, B.C., CP 22800, Mexico}
\altaffiltext{3}{South African Astronomical Observatory, PO Box 9,
                 Observatory 7935, Cape Town, South Africa}
\altaffiltext{4}{UNIFEI, DFQ -- Instituto de Ci\^encias Exatas,
                 Universidade Federal de Itajub\'a, Itajub\'a MG, Brasil}
\altaffiltext{5}{Departamento de Astronomia do IAG/USP, 
                 Universidade de S\~ao Paulo,
                 Rua do M\~atao 1226, S\~ao Paulo, 05508-900, SP, Brasil}
\altaffiltext{6}{Max Planck Institut f\"ur Astrophysik,
                 Karl-Schwarzschild-Str.\ 1,
                 Postfach 1317, D-85741 Garching, Germany}
\altaffiltext{7}{Laborat\'orio Nacional de Astrof\'isica/MCT,
                 Rua Estados Unidos 154, 37504-364 - Itajub\'a, MG, Brasil}


\begin{abstract}
Photometric data in the \ubvri\ system have been acquired for 80 solar analog stars for which we have previously derived highly precise atmospheric parameters $\teff$, $\logg$, and $\feh$ using high resolution, high signal-to-noise ratio spectra. UBV and (RI)$_\mathrm{C}$ data for 46 and 76 of these stars, respectively, are published for the first time. Combining our data with those from the literature, colors in the \ubvri\ system, with $\simeq0.01$\,mag precision, are now available for 112 solar analogs. Multiple linear regression is used to derive the solar colors from these photometric data and the spectroscopically derived $\teff$, $\logg$, and $\feh$ values. To minimize the impact of systematic errors in the model-dependent atmospheric parameters, we use only the data for the ten stars that most closely resemble our Sun, i.e., the solar twins, and derive the following solar colors: $(B-V)_\odot=0.653\pm0.005$, $(U-B)_\odot=0.166\pm0.022$, $(V-R)_\odot=0.352\pm0.007$, and $(V-I)_\odot=0.702\pm0.010$. These colors are consistent, within the 1\,$\sigma$ errors, with those derived using the entire sample of 112 solar analogs. We also derive the solar colors using the relation between spectral line-depth ratios and observed stellar colors, i.e., with a completely model-independent approach, and without restricting the analysis to solar twins. We find: $(B-V)_\odot=0.653\pm0.003$, $(U-B)_\odot=0.158\pm0.009$, $(V-R)_\odot=0.356\pm0.003$, and $(V-I)_\odot=0.701\pm0.003$, in excellent agreement with the model-dependent analysis.
\end{abstract}


\section{INTRODUCTION}

Our Sun is the primary reference in stellar astrophysics. Its fundamental parameters are known with a precision and accuracy far greater than those of any other astronomical object known. Observationally, however, comparing the Sun with the distant stars is not an easy task. Unless dedicated to solar observation, or carefully adapted for that purpose, telescopes and their instruments are designed to collect as much light as possible from faint targets. Any attempt to observe the Sun with the same instrumental setup used to observe the distant stars will suffer from saturation. Fortunately, the Sun as a star can be studied indirectly, in particular using stars that have spectral features very similar to those observed in the solar spectrum, i.e., solar analog stars \cite[e.g.,][]{cayrel96}.

A wealth of useful information on the physical properties of stars can be inferred from their photometry. Narrow band systems such as Str\"omgren's $uvby$-$\beta$ \citep{stromgren63} and systems designed for very large, all-sky surveys such as the $ugriz$ system \cite[e.g.,][]{fukugita96} are in many ways superior, or at least complementary, to the Johnson-Cousins \ubvri\ system \citep{johnson53,cousins76}. Nevertheless, for historical reasons, one could argue that the latter is still one of the most important ones \cite[e.g.,][]{bessell05:review}. Much of our knowledge on stars is based on this type of observational data, and it is no surprise that whenever a new photometric system is introduced, transformation equations to the \ubvri\ system must be determined.

Theoretical models can be used to translate photometric data into physical parameters, and vice versa. These relationships, however, must be able to reproduce very well the solar values, given the high precision and accuracy with which the solar properties are known. The problem is that the solar colors cannot be measured directly, i.e., in an identical fashion as those of the distant stars, as explained before. Since they need to be derived indirectly, they are typically very uncertain and not very useful for the calibration of stellar models. Thus the need for refinement in the derivation of the solar colors whenever possible.

The solar colors in the \ubvri\ system, in particular $(B-V)_\odot$, have been a subject of debate for many decades. Values found in the literature, as derived by many different authors using a variety of techniques, range from about 0.62 to 0.69. Using the effective temperature ($\teff$) versus $(B-V)$ relation by \cite{casagrande10}, and adopting $\feh=0$, one finds that this range of $(B-V)$ color corresponds to a $\teff$ range of 216\,K. Such large uncertainty in a fundamental zero point calibration represents a severe limitation for reliably constraining stellar models.

A few direct measurements of the $(B-V)$ solar color have been made \cite[e.g.,][]{stebbins57,tug82}, but the range of $(B-V)_\odot$ values reported is essentially the same as that corresponding to the indirect measurements, suggesting that instrumental effects are very difficult to control \cite[e.g.,][]{vandenbergh65}. Indirectly, the solar colors can be measured using samples of stars with known physical properties and interpolating the correlation between these parameters and observed colors to the solar values \cite[e.g.,][]{chmielewski81,ramirez05b,holmberg06,casagrande10}. In some cases, other types of observations, for example spectroscopic or spectrophotometric, of the Sun and the distant stars, are used, in addition to the stellar photometry, to interpolate to the solar values
 \cite[e.g.,][]{clements79,straizys94,gray92:colors}. The large range of $(B-V)_\odot$ values found in the literature (0.62--0.69), and the fact that the average error in the $(B-V)$ values typically measured with present-day instrumentation for the distant stars is only about 0.01\,mag, suggest that systematic errors are still the dominant source of uncertainty for indirect determinations of $(B-V)_\odot$. For older reviews and a complete list of references on $(B-V)_\odot$, we refer the reader to \citet[][his Table~2]{chmielewski81} and \citet[][his Figure~1]{gray92:colors}.

In a more recent revival of the $(B-V)_\odot$ debate, \cite{ramirez05a,ramirez05b} and \cite{casagrande06} have both used the so-called infrared flux method \cite[IRFM,][]{blackwell79} to derive the effective temperatures of large samples of nearby stars with accurate $\logg$ and $\feh$ values, which were then used to calibrate $\feh$-dependent $\teff$-color relations. Using the latter, interpolation to the solar $\teff=5777$\,K and $\feh=0$ allowed them to infer $(B-V)_\odot$, among other solar colors. Interestingly, even though both groups used the same technique to derive the star's $\teff$ values, their inferred solar colors differ by about 0.03\,mag. While \cite{ramirez05b} suggest a ``blue'' $(B-V)_\odot=0.619$, \cite{casagrande06} find a more ``red'' $(B-V)_\odot=0.651$. Although in principle nearly consistent within the 1\,$\sigma$ uncertainties, which are about 0.02\,mag for each, this discrepancy has been traced back to a difference in the zero point of the absolute flux calibration in the IRFM. \cite{casagrande10} have fine-tuned this absolute calibration and validated their IRFM $\teff$ scale using interferometrically measured stellar angular diameters and HST spectrophotometry. Their implementation of the IRFM gives us the most reliable $\teff$ scale available today, from which they infer $(B-V)_\odot=0.641\pm0.024$. The relatively large size of the error bar compared to the typical error in $(B-V)$ measurements ($\simeq0.01$\,mag) is due to the fact that $\teff$-color relations of a sample of stars covering a wide range of stellar parameters was used, thus propagating small, but non-negligible, systematic errors into the analysis.

In recent years, we have undertaken the task of studying solar twin and analog stars, i.e., stars with atmospheric parameters $\teff$, $\logg$, and $\feh$ identical and very similar to those of our Sun, respectively. We have carried out spectroscopic surveys in both the southern and northern hemispheres, searching for these stars and performing unprecedentedly high precision spectroscopic analysis \cite[e.g.,][]{melendez06:twins,melendez09:twins,melendez07:twins,ramirez09}. Surprisingly for us, before the present work, photometric data in the \ubvri\ system for the solar twins and analogs that we identified were scarce in the literature. For example, only about half of the stars of interest were found in the UBV section of the General Catalogue of Photometric Data \cite[GCPD,][]{mermilliod97} and the {\it Hipparcos} catalog $(B-V)$ compilation \citep{perryman97}. Motivated by this lack of fundamental, very important astronomical data, we have carried out campaigns to measure colors of solar analog stars in the \ubvri\ system at three different locations, which allowed us to cover the entire sky. In this paper, we present the photometric data acquired and use them along with our spectroscopically determined stellar atmospheric parameters, as well as the high quality spectra themselves, to derive the solar \ubvri\ colors. We expect these solar colors to be both very precise and accurate because the sample selection guarantees that the impact of systematic errors is small. For the first time, a statistically significant sample of solar twins and analogs with highly precise differential stellar parameters derived from high quality spectra, and homogeneously measured photometry, are available to derive the \ubvri\ colors of the Sun.

\begin{deluxetable*}{lcccccc}
\tablewidth{0pc}
\tablecaption{SAAO Photometry}
\tabletypesize{\tiny}
\tablehead{\colhead{HIP} & \colhead{$V$} & \colhead{$(B-V)$} & \colhead{$(U-B)$} & \colhead{$(V-R)$} & \colhead{$(V-I)$} & \colhead{$N_\mathrm{obs}$}}
\startdata
   348 & $8.602\pm0.015$ & $0.669\pm0.015$ & $0.124\pm0.015$ & $0.346\pm0.015$ & $0.691\pm0.015$ & 1 \\ 
   996 & $8.215\pm0.015$ & $0.664\pm0.015$ & $0.163\pm0.015$ & $0.352\pm0.015$ & $0.694\pm0.015$ & 1 \\ 
  1499 & $6.474\pm0.012$ & $0.687\pm0.008$ & $0.257\pm0.008$ & $0.368\pm0.005$ & $0.715\pm0.005$ & 2 \\ 
  4909 & $8.505\pm0.025$ & $0.636\pm0.006$ & $0.133\pm0.015$ & $0.363\pm0.013$ & $0.689\pm0.016$ & 2 \\ 
  5134 & $8.979\pm0.009$ & $0.640\pm0.004$ & $0.081\pm0.015$ & $0.345\pm0.008$ & $0.706\pm0.013$ & 2 \\ 
  6407 & $8.625\pm0.004$ & $0.656\pm0.004$ & $0.144\pm0.015$ & $0.360\pm0.011$ & $0.704\pm0.015$ & 2 \\ 
  8507 & $8.898\pm0.004$ & $0.651\pm0.006$ & $0.141\pm0.023$ & $0.359\pm0.011$ & $0.730\pm0.007$ & 2 \\ 
  8841 & $9.246\pm0.006$ & $0.669\pm0.004$ & $0.157\pm0.014$ & $0.378\pm0.004$ & $0.729\pm0.015$ & 2 \\ 
  9349 & $7.991\pm0.004$ & $0.650\pm0.004$ & $0.147\pm0.008$ & $0.343\pm0.004$ & $0.691\pm0.012$ & 2 \\ 
 11915 & $8.615\pm0.008$ & $0.653\pm0.004$ & $0.134\pm0.004$ & $0.354\pm0.004$ & $0.699\pm0.004$ & 2 \\ 
 28336 & $8.995\pm0.015$ & $0.642\pm0.015$ & $0.130\pm0.015$ & $0.360\pm0.015$ & $0.710\pm0.015$ & 1 \\ 
 30037 & $9.162\pm0.015$ & $0.682\pm0.015$ & $0.213\pm0.015$ & $0.361\pm0.015$ & $0.706\pm0.015$ & 1 \\ 
 30502 & $8.667\pm0.015$ & $0.664\pm0.015$ & $0.152\pm0.015$ & $0.368\pm0.015$ & $0.707\pm0.015$ & 1 \\ 
 36512 & $7.733\pm0.004$ & $0.655\pm0.005$ & $0.120\pm0.004$ & $0.355\pm0.005$ & $0.696\pm0.015$ & 2 \\ 
 38072 & $9.222\pm0.004$ & $0.648\pm0.004$ & $0.151\pm0.011$ & $0.363\pm0.006$ & $0.701\pm0.004$ & 2 \\ 
 39748 & $8.591\pm0.006$ & $0.615\pm0.011$ & $0.050\pm0.004$ & $0.340\pm0.008$ & $0.681\pm0.004$ & 2 \\ 
 41317 & $7.809\pm0.006$ & $0.664\pm0.008$ & $0.159\pm0.004$ & $0.367\pm0.004$ & $0.714\pm0.011$ & 2 \\ 
 43190 & $8.508\pm0.015$ & $0.670\pm0.015$ & $0.232\pm0.015$ & $0.370\pm0.015$ & $0.696\pm0.015$ & 1 \\ 
 44935 & $8.688\pm0.015$ & $0.654\pm0.015$ & $0.182\pm0.015$ & $0.345\pm0.015$ & $0.684\pm0.015$ & 1 \\ 
 44997 & $8.325\pm0.015$ & $0.666\pm0.015$ & $0.191\pm0.015$ & $0.344\pm0.015$ & $0.685\pm0.015$ & 1 \\ 
 46126 & $8.514\pm0.006$ & $0.653\pm0.010$ & $0.167\pm0.006$ & $0.354\pm0.006$ & $0.704\pm0.023$ & 2 \\ 
 49756 & $7.525\pm0.015$ & $0.644\pm0.015$ & $0.181\pm0.015$ & $0.349\pm0.015$ & $0.672\pm0.015$ & 1 \\ 
 51258 & $7.874\pm0.004$ & $0.730\pm0.004$ & $0.344\pm0.008$ & $0.386\pm0.006$ & $0.735\pm0.004$ & 2 \\ 
 54102 & $8.653\pm0.004$ & $0.649\pm0.004$ & $0.142\pm0.015$ & $0.346\pm0.004$ & $0.698\pm0.004$ & 2 \\ 
 55409 & $8.001\pm0.010$ & $0.657\pm0.011$ & $0.174\pm0.017$ & $0.368\pm0.007$ & $0.720\pm0.008$ & 2 \\ 
 57291 & $7.466\pm0.008$ & $0.740\pm0.004$ & $0.354\pm0.006$ & $0.375\pm0.016$ & $0.732\pm0.013$ & 2 \\ 
 59357 & $8.655\pm0.008$ & $0.627\pm0.008$ & $0.076\pm0.004$ & $0.344\pm0.004$ & $0.684\pm0.007$ & 2 \\ 
 60081 & $8.023\pm0.007$ & $0.696\pm0.006$ & $0.290\pm0.008$ & $0.373\pm0.007$ & $0.702\pm0.007$ & 2 \\ 
 60370 & $6.703\pm0.004$ & $0.651\pm0.004$ & $0.148\pm0.013$ & $0.349\pm0.008$ & $0.674\pm0.008$ & 2 \\ 
 60653 & $8.731\pm0.015$ & $0.638\pm0.015$ & $0.109\pm0.015$ & $0.358\pm0.015$ & $0.715\pm0.015$ & 1 \\ 
 64150 & $6.761\pm0.007$ & $0.688\pm0.016$ & $0.200\pm0.004$ & $0.349\pm0.017$ & $0.694\pm0.004$ & 2 \\ 
 64497 & $8.920\pm0.004$ & $0.653\pm0.004$ & $0.176\pm0.004$ & $0.357\pm0.004$ & $0.686\pm0.013$ & 2 \\ 
 64713 & $9.250\pm0.004$ & $0.648\pm0.004$ & $0.138\pm0.004$ & $0.355\pm0.010$ & $0.690\pm0.010$ & 2 \\ 
 64794 & $8.421\pm0.006$ & $0.640\pm0.006$ & $0.150\pm0.016$ & $0.343\pm0.010$ & $0.696\pm0.016$ & 2 \\ 
 64993 & $8.878\pm0.004$ & $0.650\pm0.007$ & $0.155\pm0.013$ & $0.352\pm0.006$ & $0.697\pm0.007$ & 2 \\ 
 66885 & $9.309\pm0.005$ & $0.630\pm0.012$ & $0.067\pm0.018$ & $0.366\pm0.004$ & $0.729\pm0.004$ & 2 \\ 
 69063 & $8.882\pm0.006$ & $0.632\pm0.004$ & $0.068\pm0.004$ & $0.352\pm0.004$ & $0.706\pm0.010$ & 2 \\ 
 73815 & $8.181\pm0.011$ & $0.668\pm0.008$ & $0.171\pm0.011$ & $0.360\pm0.006$ & $0.698\pm0.004$ & 2 \\ 
 74389 & $7.768\pm0.010$ & $0.640\pm0.014$ & $0.149\pm0.007$ & $0.349\pm0.004$ & $0.689\pm0.004$ & 2 \\ 
 75923 & $9.171\pm0.014$ & $0.664\pm0.006$ & $0.138\pm0.004$ & $0.367\pm0.018$ & $0.718\pm0.007$ & 2 \\ 
 77883 & $8.727\pm0.006$ & $0.681\pm0.004$ & $0.214\pm0.011$ & $0.368\pm0.004$ & $0.719\pm0.004$ & 2 \\ 
 79304 & $8.670\pm0.006$ & $0.629\pm0.007$ & $0.166\pm0.006$ & $0.353\pm0.011$ & $0.680\pm0.004$ & 2 \\ 
 79578 & $6.533\pm0.033$ & $0.678\pm0.028$ & $0.145\pm0.012$ & $0.352\pm0.008$ & $0.699\pm0.004$ & 2 \\ 
 79672 & $5.503\pm0.015$ & $0.644\pm0.015$ & $0.157\pm0.015$ & $0.354\pm0.015$ & $0.704\pm0.015$ & 1 \\ 
 82853 & $8.993\pm0.026$ & $0.660\pm0.020$ & $0.181\pm0.004$ & $0.396\pm0.004$ & $0.728\pm0.006$ & 2 \\ 
 83707 & $8.606\pm0.015$ & $0.655\pm0.004$ & $0.181\pm0.004$ & $0.348\pm0.011$ & $0.699\pm0.007$ & 2 \\ 
 85042 & $6.287\pm0.004$ & $0.669\pm0.004$ & $0.207\pm0.017$ & $0.364\pm0.020$ & $0.707\pm0.051$ & 2 \\ 
 85272 & $9.121\pm0.010$ & $0.640\pm0.011$ & $0.095\pm0.012$ & $0.368\pm0.011$ & $0.718\pm0.008$ & 2 \\ 
 85285 & $8.378\pm0.019$ & $0.632\pm0.008$ & $0.076\pm0.018$ & $0.363\pm0.019$ & $0.715\pm0.005$ & 2 \\ 
 86796 & $5.124\pm0.006$ & $0.681\pm0.028$ & $0.296\pm0.017$ & $0.386\pm0.005$ & $0.706\pm0.005$ & 2 \\ 
 89162 & $8.903\pm0.007$ & $0.658\pm0.005$ & $0.176\pm0.004$ & $0.363\pm0.010$ & $0.698\pm0.011$ & 2 \\ 
 89650 & $8.943\pm0.010$ & $0.644\pm0.006$ & $0.126\pm0.004$ & $0.354\pm0.010$ & $0.679\pm0.013$ & 2 \\ 
 91332 & $7.971\pm0.008$ & $0.696\pm0.009$ & $0.263\pm0.004$ & $0.365\pm0.015$ & $0.705\pm0.004$ & 2 \\ 
102152 & $9.188\pm0.013$ & $0.671\pm0.018$ & $0.196\pm0.023$ & $0.382\pm0.004$ & $0.727\pm0.004$ & 2 \\ 
104504 & $8.542\pm0.015$ & $0.640\pm0.015$ & $0.081\pm0.015$ & $0.366\pm0.015$ & $0.696\pm0.015$ & 1 \\ 
108996 & $8.856\pm0.015$ & $0.640\pm0.015$ & $0.165\pm0.015$ & $0.350\pm0.015$ & $0.677\pm0.015$ & 1 \\ 
118159 & $9.017\pm0.004$ & $0.633\pm0.018$ & $0.090\pm0.015$ & $0.355\pm0.014$ & $0.679\pm0.005$ & 2 

\enddata
\label{t:saao}
\end{deluxetable*}

\begin{deluxetable*}{lcccccc}
\tablewidth{0pc}
\tablecaption{SPM Photometry}
\tabletypesize{\tiny}
\tablehead{\colhead{HIP} & \colhead{$V$} & \colhead{$(B-V)$} & \colhead{$(U-B)$} & \colhead{$(V-R)$} & \colhead{$(V-I)$} & \colhead{$N_\mathrm{obs}$}}
\startdata
   348 & $8.595\pm0.020$ & $0.636\pm0.022$ & $0.095\pm0.021$ & $0.376\pm0.036$ & $0.706\pm0.022$ & 1 \\ 
   996 & $8.189\pm0.009$ & $0.630\pm0.012$ & $0.114\pm0.021$ & $0.372\pm0.011$ & $0.689\pm0.011$ & 1 \\ 
  2131 & $8.923\pm0.008$ & $0.642\pm0.011$ & $0.106\pm0.023$ & $0.376\pm0.009$ & $0.720\pm0.009$ & 1 \\ 
  2894 & $8.651\pm0.018$ & $0.659\pm0.028$ & $0.200\pm0.039$ & $0.371\pm0.034$ & $0.703\pm0.019$ & 1 \\ 
  4909 & $8.515\pm0.009$ & $0.633\pm0.011$ & $0.106\pm0.014$ & $0.372\pm0.012$ & $0.688\pm0.013$ & 1 \\ 
  5134 & $8.969\pm0.007$ & $0.624\pm0.009$ & $0.066\pm0.017$ & $0.365\pm0.010$ & $0.702\pm0.008$ & 1 \\ 
  6407 & $8.613\pm0.021$ & $0.649\pm0.022$ & $0.124\pm0.017$ & $0.370\pm0.035$ & $0.704\pm0.021$ & 1 \\ 
  7245 & $8.361\pm0.009$ & $0.667\pm0.013$ & $0.186\pm0.022$ & $0.376\pm0.011$ & $0.691\pm0.011$ & 1 \\ 
  8507 & \nodata & \nodata & $0.126\pm0.014$ & \nodata & \nodata & 1 \\ 
  9349 & $8.220\pm0.054$ & \nodata & \nodata & \nodata & \nodata & 1 \\ 
 18261 & $7.980\pm0.027$ & $0.616\pm0.029$ & $0.083\pm0.017$ & $0.348\pm0.047$ & $0.670\pm0.028$ & 1 \\ 
 25670 & $8.275\pm0.021$ & $0.663\pm0.023$ & $0.167\pm0.012$ & $0.362\pm0.036$ & $0.698\pm0.024$ & 1 \\ 
 28336 & $8.998\pm0.005$ & $0.647\pm0.013$ & $0.091\pm0.013$ & $0.366\pm0.017$ & $0.687\pm0.026$ & 1 \\ 
 36512 & $7.700\pm0.011$ & $0.668\pm0.019$ & $0.134\pm0.019$ & $0.406\pm0.029$ & $0.685\pm0.099$ & 1 \\ 
 38072 & $9.214\pm0.016$ & $0.660\pm0.025$ & $0.125\pm0.027$ & $0.362\pm0.026$ & $0.693\pm0.021$ & 2 \\ 
 41317 & $7.798\pm0.015$ & $0.673\pm0.023$ & $0.127\pm0.029$ & $0.386\pm0.034$ & $0.730\pm0.024$ & 1 \\ 
 44324 & $7.943\pm0.010$ & $0.620\pm0.027$ & $0.083\pm0.035$ & $0.342\pm0.011$ & $0.674\pm0.016$ & 4 \\ 
 44935 & $8.743\pm0.004$ & $0.643\pm0.006$ & $0.203\pm0.006$ & $0.364\pm0.006$ & $0.691\pm0.006$ & 1 \\ 
 44997 & $8.370\pm0.015$ & $0.650\pm0.016$ & $0.204\pm0.007$ & $0.383\pm0.027$ & $0.713\pm0.016$ & 1 \\ 
 46066 & $8.928\pm0.007$ & $0.664\pm0.017$ & $0.188\pm0.016$ & $0.381\pm0.008$ & $0.717\pm0.008$ & 2 \\ 
 49572 & $9.288\pm0.006$ & $0.640\pm0.007$ & $0.138\pm0.006$ & $0.357\pm0.008$ & $0.702\pm0.007$ & 1 \\ 
 49756 & $7.540\pm0.004$ & $0.647\pm0.006$ & $0.186\pm0.006$ & $0.361\pm0.006$ & $0.691\pm0.006$ & 1 \\ 
 55459 & $7.646\pm0.004$ & $0.646\pm0.006$ & $0.153\pm0.006$ & $0.359\pm0.006$ & $0.692\pm0.006$ & 1 \\ 
 56948 & $8.669\pm0.004$ & $0.646\pm0.006$ & $0.180\pm0.006$ & $0.360\pm0.006$ & $0.680\pm0.006$ & 1 \\ 
 59357 & $8.752\pm0.004$ & $0.662\pm0.010$ & $0.090\pm0.011$ & $0.388\pm0.006$ & $0.746\pm0.006$ & 1 \\ 
 60314 & $8.780\pm0.008$ & $0.665\pm0.009$ & $0.155\pm0.007$ & $0.358\pm0.011$ & $0.676\pm0.009$ & 1 \\ 
 62175 & $8.011\pm0.005$ & $0.656\pm0.006$ & $0.194\pm0.007$ & $0.366\pm0.006$ & $0.682\pm0.006$ & 1 \\ 
 64150 & $6.883\pm0.007$ & $0.717\pm0.008$ & $0.217\pm0.006$ & $0.402\pm0.009$ & $0.724\pm0.008$ & 1 \\ 
 64497 & $9.035\pm0.004$ & $0.701\pm0.006$ & $0.194\pm0.006$ & $0.397\pm0.006$ & \nodata & 1 \\ 
 64713 & $9.297\pm0.008$ & $0.669\pm0.020$ & $0.187\pm0.019$ & $0.367\pm0.012$ & $0.727\pm0.009$ & 1 \\ 
 64794 & $8.461\pm0.020$ & $0.667\pm0.020$ & $0.143\pm0.006$ & $0.377\pm0.030$ & $0.710\pm0.020$ & 1 \\ 
 64993 & $8.921\pm0.004$ & $0.666\pm0.009$ & $0.172\pm0.009$ & $0.365\pm0.006$ & $0.712\pm0.006$ & 1 \\ 
 66885 & $9.274\pm0.010$ & $0.628\pm0.023$ & $0.096\pm0.024$ & $0.353\pm0.016$ & $0.741\pm0.011$ & 1 \\ 
 73815 & $8.173\pm0.005$ & $0.668\pm0.006$ & $0.161\pm0.008$ & $0.363\pm0.006$ & $0.683\pm0.009$ & 2 \\ 
 74341 & $8.857\pm0.005$ & $0.673\pm0.014$ & $0.165\pm0.018$ & $0.354\pm0.016$ & $0.684\pm0.009$ & 3 \\ 
 74389 & $7.760\pm0.021$ & $0.623\pm0.024$ & $0.202\pm0.024$ & $0.352\pm0.022$ & $0.667\pm0.023$ & 1 \\ 
 75923 & $9.149\pm0.005$ & $0.651\pm0.006$ & $0.134\pm0.006$ & $0.363\pm0.007$ & $0.689\pm0.006$ & 1 \\ 
 77883 & $8.770\pm0.004$ & $0.700\pm0.006$ & $0.227\pm0.006$ & $0.395\pm0.006$ & $0.751\pm0.006$ & 1 \\ 
 78028 & $8.651\pm0.012$ & $0.638\pm0.019$ & $0.118\pm0.024$ & $0.355\pm0.016$ & $0.683\pm0.016$ & 5 \\ 
 78680 & $8.243\pm0.013$ & $0.626\pm0.018$ & $0.079\pm0.021$ & $0.358\pm0.016$ & $0.698\pm0.016$ & 3 \\ 
 79186 & $8.341\pm0.014$ & $0.675\pm0.028$ & $0.140\pm0.033$ & $0.377\pm0.022$ & $0.724\pm0.018$ & 3 \\ 
 79304 & $8.718\pm0.004$ & $0.661\pm0.007$ & $0.148\pm0.008$ & $0.365\pm0.006$ & $0.705\pm0.011$ & 2 \\ 
 79672 & $5.522\pm0.019$ & $0.680\pm0.019$ & $0.182\pm0.006$ & $0.401\pm0.019$ & $0.773\pm0.019$ & 2 \\ 
 81512 & $9.245\pm0.015$ & $0.652\pm0.017$ & $0.140\pm0.019$ & $0.374\pm0.019$ & $0.712\pm0.016$ & 3 \\ 
 85285 & $8.356\pm0.010$ & $0.642\pm0.020$ & $0.049\pm0.021$ & $0.362\pm0.013$ & $0.689\pm0.017$ & 3 \\ 
 88194 & $7.084\pm0.011$ & $0.656\pm0.012$ & $0.132\pm0.016$ & $0.390\pm0.016$ & $0.723\pm0.018$ & 3 \\ 
 88427 & $9.329\pm0.006$ & $0.638\pm0.013$ & $0.089\pm0.022$ & $0.357\pm0.018$ & $0.704\pm0.007$ & 1 \\ 
 89443 & $8.843\pm0.004$ & $0.660\pm0.007$ & $0.147\pm0.013$ & $0.380\pm0.006$ & $0.715\pm0.006$ & 1 \\ 
100963 & $7.081\pm0.009$ & $0.651\pm0.014$ & $0.128\pm0.024$ & $0.359\pm0.010$ & $0.708\pm0.010$ & 1 \\ 
102152 & $9.220\pm0.010$ & $0.667\pm0.017$ & $0.158\pm0.022$ & $0.383\pm0.032$ & $0.715\pm0.016$ & 1 \\ 
104504 & $8.594\pm0.020$ & $0.636\pm0.022$ & $0.057\pm0.014$ & $0.361\pm0.039$ & $0.703\pm0.021$ & 1 \\ 
108708 & $8.945\pm0.017$ & $0.659\pm0.020$ & $0.162\pm0.012$ & $0.379\pm0.036$ & $0.707\pm0.021$ & 1 \\ 
108996 & $8.889\pm0.009$ & $0.659\pm0.010$ & $0.103\pm0.029$ & $0.357\pm0.010$ & $0.688\pm0.014$ & 2 \\ 
109931 & $8.956\pm0.019$ & $0.674\pm0.020$ & $0.204\pm0.020$ & $0.388\pm0.034$ & $0.710\pm0.020$ & 1 \\ 
118159 & $9.004\pm0.004$ & $0.627\pm0.007$ & $0.090\pm0.014$ & $0.358\pm0.007$ & $0.681\pm0.006$ & 1 

\enddata
\label{t:spm}
\end{deluxetable*}

\begin{deluxetable*}{lccccc}
\tablewidth{0pc}
\tablecaption{OPD Photometry}
\tabletypesize{\tiny}
\tablehead{\colhead{HIP} & \colhead{$V$} & \colhead{$(B-V)$} & \colhead{$(V-R)$} & \colhead{$(V-I)$} & \colhead{$N_\mathrm{obs}$}}
\startdata
   348 & $8.604\pm0.026$ & \nodata & $0.348\pm0.017$ & $0.688\pm0.035$ & 1 \\ 
   996 & $8.216\pm0.028$ & \nodata & $0.347\pm0.018$ & $0.666\pm0.037$ & 1 \\ 
  4909 & $8.498\pm0.024$ & $0.647\pm0.015$ & $0.356\pm0.016$ & $0.674\pm0.032$ & 1 \\ 
  5134 & $8.970\pm0.025$ & $0.623\pm0.016$ & $0.352\pm0.016$ & $0.702\pm0.034$ & 1 \\ 
  6407 & $8.617\pm0.025$ & $0.633\pm0.016$ & $0.361\pm0.016$ & $0.705\pm0.034$ & 1 \\ 
  7245 & $8.366\pm0.032$ & \nodata & $0.350\pm0.021$ & $0.685\pm0.043$ & 1 \\ 
  8507 & $8.924\pm0.024$ & $0.654\pm0.016$ & $0.370\pm0.016$ & $0.721\pm0.033$ & 1 \\ 
 39748 & $8.675\pm0.058$ & $0.614\pm0.067$ & \nodata & \nodata & 2 \\ 
 49756 & $7.577\pm0.020$ & \nodata & $0.347\pm0.010$ & $0.678\pm0.012$ & 2 \\ 
 55409 & $8.149\pm0.102$ & \nodata & $0.366\pm0.020$ & $0.713\pm0.025$ & 3 \\ 
 59357 & $8.676\pm0.020$ & \nodata & $0.353\pm0.009$ & $0.682\pm0.012$ & 3 \\ 
 60653 & $8.742\pm0.020$ & \nodata & $0.361\pm0.009$ & $0.694\pm0.012$ & 3 \\ 
 64497 & $8.995\pm0.020$ & \nodata & $0.357\pm0.009$ & $0.688\pm0.012$ & 3 \\ 
 64713 & $9.429\pm0.057$ & $0.674\pm0.065$ & \nodata & \nodata & 2 \\ 
 64794 & $8.867\pm0.064$ & \nodata & \nodata & \nodata & 3 \\ 
 73815 & $8.137\pm0.108$ & \nodata & $0.364\pm0.021$ & $0.708\pm0.026$ & 1 \\ 
 74341 & $8.917\pm0.022$ & \nodata & $0.355\pm0.011$ & $0.699\pm0.013$ & 3 \\ 
 74389 & $7.803\pm0.020$ & \nodata & $0.349\pm0.009$ & $0.678\pm0.012$ & 3 \\ 
 75923 & $9.182\pm0.012$ & $0.658\pm0.011$ & $0.368\pm0.009$ & $0.725\pm0.014$ & 3 \\ 
 77883 & $8.734\pm0.011$ & $0.691\pm0.011$ & $0.377\pm0.009$ & $0.738\pm0.013$ & 3 \\ 
 79304 & $8.703\pm0.013$ & $0.656\pm0.014$ & $0.351\pm0.011$ & $0.692\pm0.016$ & 3 \\ 
 82853 & $8.018\pm0.151$ & \nodata & $0.376\pm0.031$ & $0.779\pm0.038$ & 1 \\ 
 83707 & $8.528\pm0.108$ & \nodata & $0.358\pm0.021$ & $0.699\pm0.026$ & 1 \\ 
 85272 & $9.120\pm0.030$ & $0.600\pm0.022$ & $0.347\pm0.019$ & $0.689\pm0.039$ & 1 \\ 
 85285 & $8.400\pm0.025$ & $0.606\pm0.017$ & $0.347\pm0.017$ & $0.682\pm0.034$ & 2 \\ 
 88194 & $7.171\pm0.022$ & \nodata & $0.356\pm0.011$ & $0.706\pm0.014$ & 2 \\ 
 89162 & $8.882\pm0.031$ & \nodata & $0.347\pm0.020$ & $0.675\pm0.041$ & 2 \\ 
 89650 & $8.946\pm0.011$ & $0.641\pm0.010$ & $0.357\pm0.009$ & $0.697\pm0.013$ & 5 \\ 
100963 & $7.140\pm0.022$ & \nodata & $0.346\pm0.011$ & $0.694\pm0.014$ & 2 \\ 
104504 & $8.532\pm0.010$ & $0.617\pm0.008$ & $0.363\pm0.008$ & $0.703\pm0.012$ & 3 \\ 
108708 & $8.937\pm0.010$ & $0.659\pm0.008$ & $0.368\pm0.008$ & $0.712\pm0.012$ & 3 \\ 
108996 & $8.881\pm0.010$ & $0.643\pm0.008$ & $0.360\pm0.008$ & $0.703\pm0.012$ & 3 \\ 
118159 & $9.005\pm0.025$ & $0.623\pm0.016$ & $0.344\pm0.016$ & $0.667\pm0.033$ & 1 

\enddata
\label{t:lna}
\end{deluxetable*}

\section{SAMPLE AND PHOTOMETRIC DATA} \label{s:sample}

The stars used in this work are listed in Table~4 of \cite{baumann10}, who studied the evolution of lithium abundances in Sun-like stars using high resolution, high signal-to-noise ratio spectra acquired by \cite{ramirez09} and \cite{melendez09:twins}. These spectra were taken using the R.\,G.\,Tull coud\'e spectrograph on the 2.7\,m Telescope at McDonald Observatory and the MIKE spectrograph on the 6.5\,m Clay/Magellan Telescope at Las Campanas Observatory. The spectral resolution $R=\lambda/\Delta\lambda$ of the spectroscopic data is about 60,000 while the signal-to-noise ratios range from about 150 to 600, with a median value closer to 400. The stellar parameters $\teff$, $\logg$, and $\feh$ used in this work are those listed in \cite{baumann10} and they were determined by forcing excitation/ionization equilibrium of iron lines in the stellar spectra. Given the high quality of the data and the careful sample selection, the average errors in the stellar parameters are only $\Delta\teff=41$\,K, $\Delta\logg=0.06$, and $\Delta\feh=0.03$, although they are significantly smaller for the stars that are most similar to the Sun. Systematic errors are not included in these error estimates, but we expect them to be very small because of the strictly differential approach we used to derive the atmospheric parameters. All of the objects analyzed in the present study are main-sequence stars, as confirmed by their $\logg$ values. We refer the reader to \cite{melendez09:twins}, \cite{ramirez09}, and \cite{baumann10} for details on the spectroscopic data reduction, the determination of stellar parameters, and the assessment of errors.

\ubvri\ magnitudes and colors for as many as possible of the stars in \cite{baumann10} were measured at three sites: SAAO (South African Astronomical Observatory), SPM (San Pedro Martir, in M\'exico), and OPD (Observat\'orio do Pico dos Dias, in Brazil); 57 stars were observed at SAAO, 55 at SPM, and 33 at OPD. A number of stars were observed at more than one location; the total number of unique stars observed is 80. Below we describe briefly our photometric observations.

The SAAO \ubvri\ observations were made using the 0.5\,m telescope and a photomultiplier tube (PMT) based modular photometer at Sutherland \citep{kilkenny88}. The PMT is a Hamamatsu R943-02 Gallium Arsenide tube and it is thermoelectrically cooled to low temperatures to reduce dark counts to minimum levels. Observations were carried out throughout 2010 and 2011 in blocks of several weeks spread over the two years. Observations of both the target objects and E-region standard stars \cite[e.g.,][]{menzies89} were made each night through the \ubvri\ filters, mostly alternating between a standard and target objects. The observations were later corrected to the \ubvri\ system using nightly observations of the E-region standards and current transformation equations that are maintained and regularly updated (about twice a year) at the Observatory. 
Observations were done only during photometric nights, and any standard star observations that deviate from the standard magnitude by more than $\pm0.05$ is not used in the reduction or determination of zero points for transformation to the \ubvri\ system. 
Based on the observations of standard stars made in multiple observing nights and/or runs, we estimate an accuracy of about 0.01\,mag for the SAAO measurements of visual magnitudes and colors. The E-region stars used in our reductions have visual magnitudes from about $V=5$ to $V=10$, which is similar to the magnitude range of our observed program stars.

The SPM observations were carried out during two runs; eight nights in May 2011 (from the 21st to the 28th) and five nights in October 2011 (from the 20th to the 24th). The San Pedro Martir 0.84\,m telescope was used, along with the Mexman filter wheel. During the May run a SITe CCD was used (1024x1024 pixels, gain=4.8 e$^-$/ADU, readout noise=13e$^-$) while in October an e2v-4290 CCD was used (4.5Kx2K pixels, gain=1.7 e$^-$/ADU, readout noise=3.8e$^-$). Sky flat fields were taken at the beginning and end of each night, and bias frames were taken between each observed field. Landolt standards were observed both at the meridian and at large air masses. All the images were bias subtracted and flat field corrected. Cosmic rays were removed using the L.A. Cosmic task \citep{vandokkum01}. Instrumental magnitudes were calculated using the IRAF \verb"photcal" package and the observations of the standard stars.

The OPD photometric data were acquired using the Zeiss 0.6\,m telescope at Pico dos Dias Observatory, operated by the Laborat\'orio Nacional de Astrof\'isica, in Brazil, during the years 2009 and 2010. The instrument used for the observation was the FOTRAP \cite[``rapid photometer'',][]{jablonski94}, which consists of a wheel with 6 filters (Johnson-Cousins \ubvri\ and clear) running at 20\,Hz and acquiring data almost simultaneously in all filters. Light from the telescope passes through the filter wheel and then by a set of diaphragms that is used for limiting interference of light from the sky and/or nearby objects. Then the light reaches the Hamamatsu photomultiplier operating at $-25$ degrees Celsius. 
Throughout the night, various \cite{graham82} standard stars are observed. Usually, one in every three stars observed was a standard, making sure no star with an airmass greater than 1.5 was observed, following the suggestion by \cite{harris81}, whose photometric reduction method is used in the reduction software of this instrument. The reduction is made using the software ``mags.exe,'' which was specially written for the instrument FOTRAP, as described in \cite{jablonski94}.

\begin{figure}
\epsscale{1.2}
\plotone{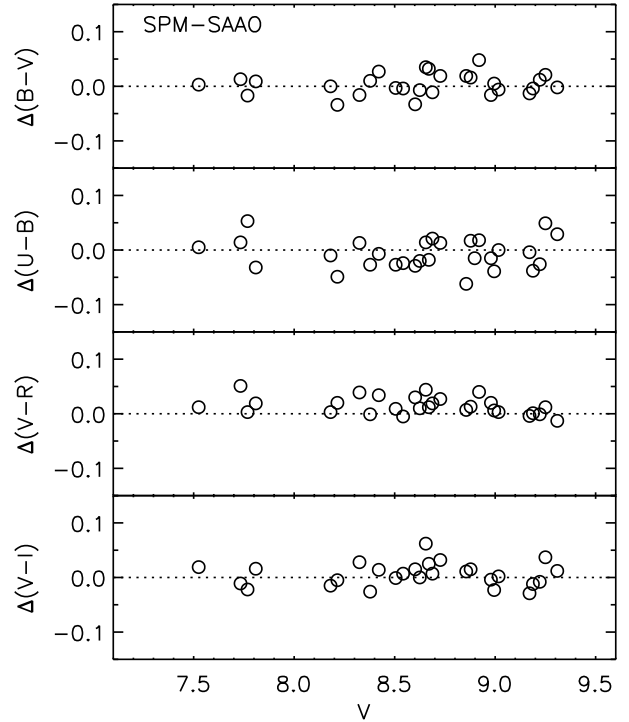}
\caption{Difference in color measured at the SPM and SAAO observatories as a function of apparent visual magnitude, as observed from SAAO.}
\label{f:spm}
\end{figure}

\begin{figure}
\epsscale{1.2}
\plotone{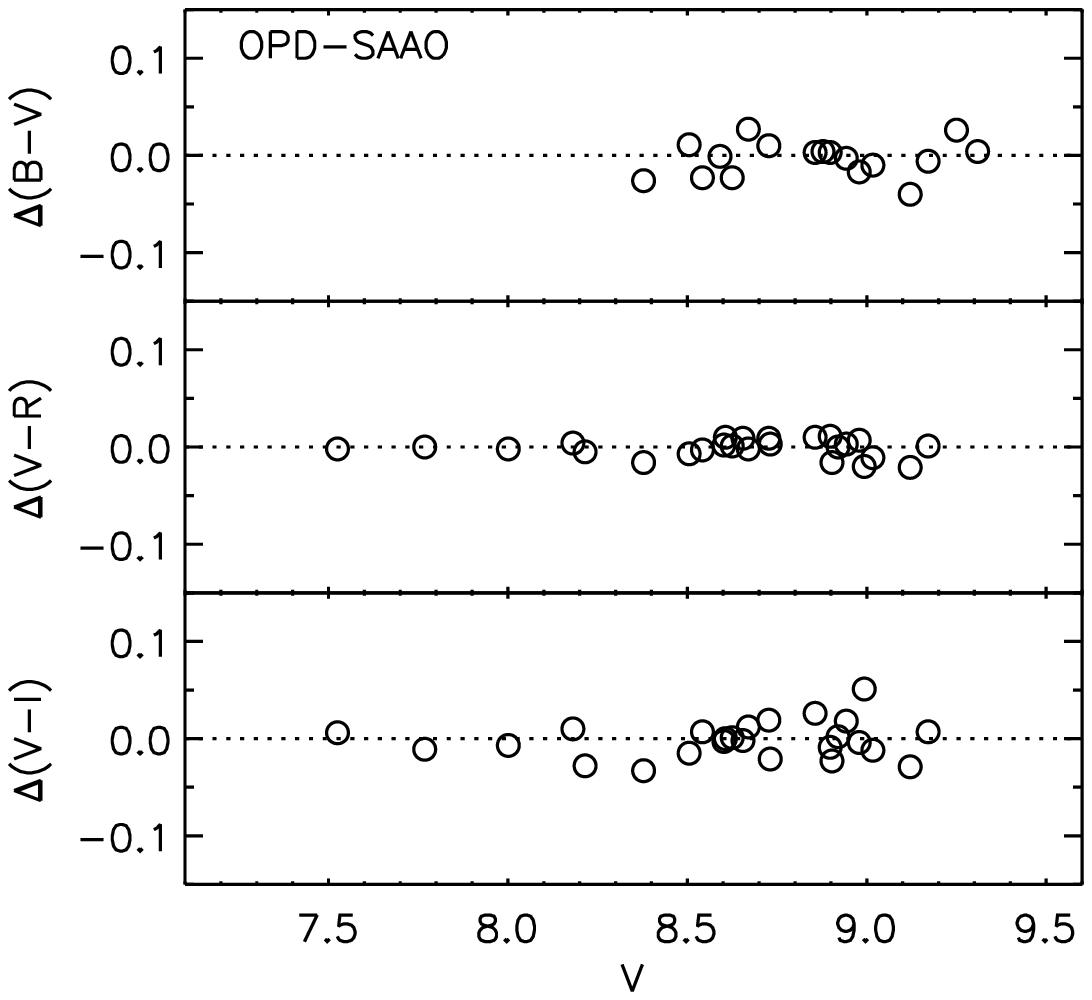}
\caption{Difference in color measured at the OPD and SAAO observatories as a function of apparent visual magnitude, as observed from SAAO.}
\label{f:lna}
\end{figure}

\begin{deluxetable}{cccc}
\tablewidth{0pc}
\tablecaption{Photometry Offsets}
\tabletypesize{\footnotesize}
\tablehead{\colhead{$\Delta(\mathrm{color})$} & \colhead{mean} & \colhead{$\sigma$} & \colhead{$n$\tablenotemark{1}} \\ \hline\hline \multicolumn{4}{c}{SPM--SAAO}}
\startdata
$\Delta(B-V)$ &  0.011 & 0.022 & 28 \\
$\Delta(U-B)$ &  0.001 & 0.021 & 29 \\
$\Delta(V-R)$ &  0.021 & 0.016 & 28 \\
$\Delta(V-I)$ &  0.011 & 0.024 & 27 \\
\cutinhead{OPD--SAAO}
$\Delta(B-V)$ & -0.004 & 0.016 & 17 \\
$\Delta(V-R)$ &  0.001 & 0.007 & 25 \\
$\Delta(V-I)$ &  0.002 & 0.015 & 25 \\
\cutinhead{TW--LIT}
$\Delta(B-V)$ &  0.002 & 0.011 & 34 \\
$\Delta(U-B)$ &  0.013 & 0.034 & 14 \\
$\Delta(V-R)$ &  0.007 & 0.006 &  4 \\
$\Delta(V-I)$ &  0.006 & 0.013 &  4

\enddata
\tablenotetext{1}{$n$ is the number of stars in common between the two samples.}
\label{t:offsets}
\end{deluxetable}

\begin{figure}
\epsscale{1.2}
\plotone{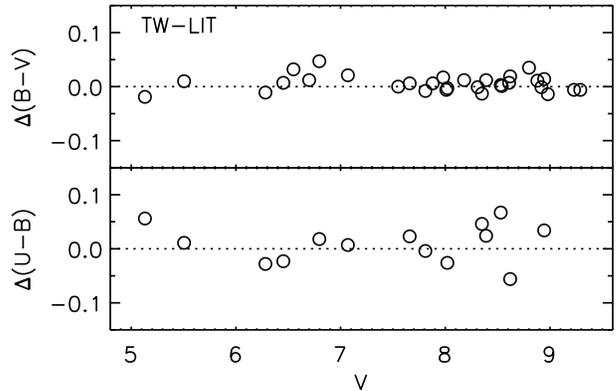}
\caption{Difference in color measured by us and those found in the literature a function of apparent visual magnitude, as found in the literature.}
\label{f:lit}
\end{figure}

The photometric data collected at the three sites described above are given in Tables~\ref{t:saao} to \ref{t:lna}. Figure~\ref{f:spm} shows the comparison of colors measured at the SPM and SAAO observatories for the stars in common. Similarly, Figure~\ref{f:lna} shows the comparison of OPD and SAAO data. In Table~\ref{t:offsets} we list the mean offsets and the star-to-star standard deviation of the difference between colors measured at different sites, determined using data for stars in common between the different samples. In most cases the mean differences are compatible with zero within the 1\,$\sigma$ uncertainties, suggesting that any offsets that could be a product of employing different sets of photometric standard stars and/or data reduction procedures are smaller than the observational errors. The only exception is the SPM $(V-R)$ data set, which shows a non-zero mean offset of $0.021\pm0.016$ relative to the SAAO data. To prevent this offset from introducing unwanted noise in our solar colors analysis, we corrected the SPM $(V-R)$ colors so that their mean difference with the SAAO data is exactly zero. The values listed in Table~\ref{t:spm}, however, are the original ones.

For the stars that were observed from more than one location, we adopted a weighted mean of the colors given from each site. The error associated to these average colors corresponds to the sample variance. However, we adopted a minimum photometric error of 0.004\,mag to prevent unreasonably small errors arising from numerical artifacts, i.e., from coincidental agreement between the (statistically few) mean values reported from different sites.

We also searched for \ubvri\ photometry in the GCPD \citep{mermilliod97} and $(B-V)$ colors in the {\it Hipparcos} catalog \citep{perryman97} for the stars in \cite{baumann10}. We used the latter only if not available in the GCDP. These {\it Hipparcos} $(B-V)$ colors correspond to those compiled by the mission team from previously published standard UBV system data sets (i.e., those with flag G in column 39 of the {\it Hipparcos} catalog), and not to the colors inferred from transformation equations using Tycho photometry (flag T instead). Sixty-six (66) stars were found with either UBV and/or RI$_\mathrm{(C)}$ colors previously reported in the literature. Thirty four (34) of these stars were observed by us. However, only four of them have RI$_\mathrm{(C)}$ data in the literature. Thus, a proper comparison of our measured colors with previously published values can only be done for $(B-V)$ and $(U-B)$. This comparison is shown in Figure~\ref{f:lit}.

The average difference in $(B-V)$ color between our measurements and those found in the literature is $\Delta(B-V)=0.002\pm0.011$, i.e., consistent with zero within the 1\,$\sigma$ uncertainty. Moreover, the star-to-star scatter in this comparison (0.011\,mag) suggests that the mean error in the measurements of $(B-V)$ is about $0.008$\,mag$=0.011/\sqrt{2}$, which is identical to the average $(B-V)$ error given in our Table~\ref{t:adopted}. Thus, our $(B-V)$ error estimates appear to be very reliable. For $(U-B)$, we derive $\Delta(U-B)=0.013\pm0.034$, also consistent with zero within the uncertainties. We also computed offsets for $(V-R)$ and $(V-I)$, but they are based on data for only four stars in common, and are therefore not so reliable. In any case, this comparison suggests that our color measurements are consistent with the \ubvri\ color scales found in the literature, and therefore with the historically adopted photometric zero points. The color offsets between our data (TW) and previously published values (LIT) are given in the lower section of Table~\ref{t:offsets}.

\begin{figure*}
\epsscale{1.19}
\plotone{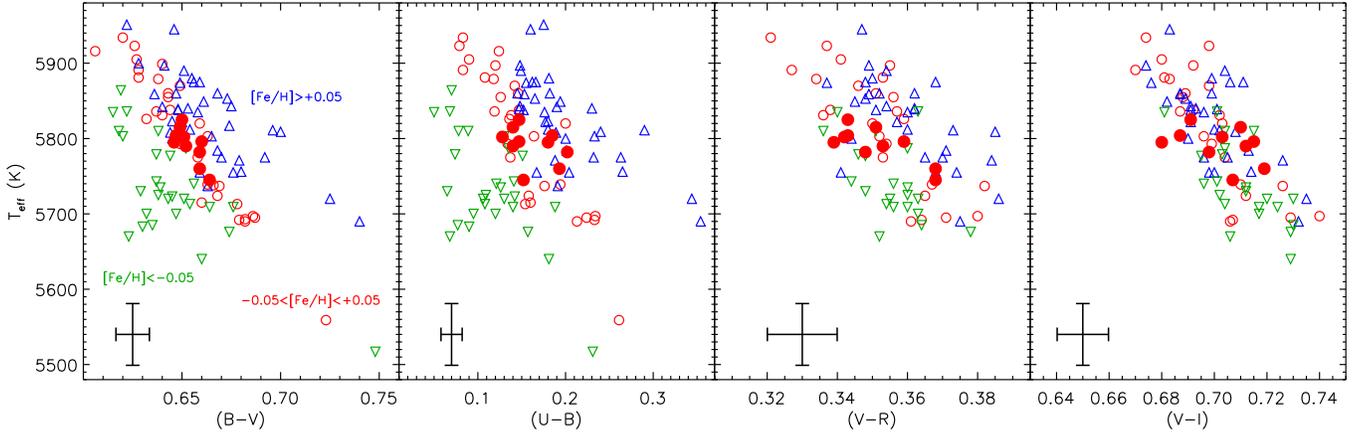}
\caption{Effective temperature versus color relations for our sample of solar twins and analogs. Open circles represent stars with near solar metallicity ($-0.05<\feh<+0.05$). Upside-down and regular triangles correspond to stars with $\feh<-0.05$ and $\feh>+0.05$, respectively. Our sample of solar twin stars is shown with filled circles. Average error bars are shown at bottom left of each panel.}
\label{f:teffcolor}
\end{figure*}

The stellar parameters and photometry adopted in this work are given in Table~\ref{t:adopted}. Here we have combined our photometric data with those found in the literature, giving equal weight to each when available for the same star. Objects for which photometric data are published for the first time are assigned mean values and errors from our measurements only. Stars not observed photometrically by us, but found in the literature, are assigned those previously published values. The average errors, in mag, of the measured colors given in Table~\ref{t:adopted} are $\Delta(B-V)=0.008$, $\Delta(U-B)=0.012$, $\Delta(V-R)=0.010$, and $\Delta(V-I)=0.010$.

\section{THE SOLAR COLORS}

\subsection{Color-$\teff$ Relations} \label{s:method1}

As is well known, colors are good indicators of $\teff$, although in many cases they can also be sensitive to other stellar parameters. Figure~\ref{f:teffcolor}, for example, shows the relation between our \ubvri\ colors and $\teff$, which clearly reveals a dependence on a second parameter, namely $\feh$, although this is much more pronounced for $(B-V)$ than $(V-I)$. The sensitivity of the \ubvri\ colors to $\logg$ is very weak, as will be shown quantitatively later in this section. As noted in Section~\ref{s:sample}, the $\teff$, $\logg$, and $\feh$ values we used were derived from our high quality spectra, using the standard excitation/ionization equilibrium balance condition for \fei\ and \feii\ lines \cite[e.g.,][]{ramirez09}. Although this technique is heavily model-dependent, the fact that our sample stars are all very similar to our Sun allows us to minimize the impact of systematic errors, because they are nearly the same for all of these objects, and because we employ a strictly differential analysis using the solar spectrum as reference.

We used the data from Table~\ref{t:adopted} to perform a multiple linear regression of the following form:
\begin{equation}
\mathrm{color}=a_0+a_1(\teff-5777\,\mathrm{K})+a_1(\logg-4.44)+a_2\feh\ ,
\label{eq:regress}
\end{equation}
from which the solar colors are inferred: $\mathrm{color}_\odot=a_0$. The error in the solar color is derived by adding in quadrature the 1\,$\sigma$ scatter of the regression, which takes into account the errors in the observed stellar colors, and the error due to the stellar parameter uncertainties. To calculate the error due to $\teff$, $\logg$, and $\feh$ uncertainties, we computed 5000 solar colors using $\teff$, $\logg$, and $\feh$ values modified randomly from their mean values, assuming a Gaussian distribution for each of the three stellar atmospheric parameters. The individual errors in these parameters for each star, as listed in Table~\ref{t:adopted}, were adopted as the standard deviations of these distributions. The 1\,$\sigma$ scatter from the 5000 tests described above was finally adopted as the uncertainty due to errors in the stellar parameters.

Two sets of solar colors were derived, a first one using the entire sample of 112 solar analogs, and a second set inferred using only the data for the 10 stars that most closely resemble the Sun (hereafter referred to as the solar twins). The solar twin sample was defined as those stars having their stellar parameters $\teff$, $\logg$, and $\feh$ within 1.4\,$\sigma$ from the solar values, where $\sigma$ is the average error in the stellar parameters of the sample. Modifying slightly the definition of solar twin star has little impact on our results. The multiplicative factor of 1.4 was chosen arbitrarily so that the sample of solar twins could have exactly ten elements.

\setcounter{table}{5}

\begin{deluxetable}{ccc}
\tablewidth{0pc}
\tabletypesize{\small}
\tablecaption{Solar Colors Inferred from $\teff$ and $\feh$ Measurements}
\tablehead{\colhead{color} & \colhead{solar twins} & \colhead{solar analogs}}
\startdata
$(B-V)$ & $0.653\pm0.005$ & $0.658\pm0.014$ \\
$(U-B)$ & $0.166\pm0.022$ & $0.163\pm0.026$ \\
$(V-R)$ & $0.352\pm0.007$ & $0.361\pm0.011$ \\
$(V-I)$ & $0.702\pm0.010$ & $0.707\pm0.012$

\enddata
\label{t:solar_colors}
\end{deluxetable}

The solar colors derived using the method described above are listed in Table~\ref{t:solar_colors}. In particular, we find $(B-V)_\odot=0.653\pm0.005$ using only the solar twin data. This value is consistent within the $1\,\sigma$ errors with that derived using the entire sample of solar analogs, $(B-V)_\odot=0.658\pm0.014$. We note, however, that the mean $(B-V)_\odot$ value increases by 0.005\,mag when using the full sample, which suggests that the effective temperatures of non solar twin stars may be slightly overestimated, making the Sun appear redder than it actually is. We find that the mean $(B-V)_\odot$ value obtained using only solar twins would be in perfect agreement with that derived using the full sample if the $\teff$ values of non solar twin stars were cooler by about 20\,K. This implies that systematic errors in the model-dependent determination of stellar parameters from iron line analysis (excitation/ionization equilibrium), although small, are non-negligible for solar analogs, but not so important for the solar twins. This is of course true only when dealing with very high quality spectroscopic data such as those used by \cite{melendez09:twins}, \cite{ramirez09}, and \cite{baumann10}, where effective temperatures with a precision comparable to 20\,K are possible to achieve.

The $(U-B)_\odot$ color has the largest error of all \ubvri\ solar colors derived; it is greater than 0.02\,mag. This is not at all surprising because $U$-filter observations and their standardization are known to be very challenging. For $(V-R)$ and $(V-I)$ we derive solar colors with errors below or about 0.01\,mag. As with $(B-V)$, the solar $(V-R)$ and $(V-I)$ colors inferred using solar twins are slightly bluer than those obtained using the full sample of solar analogs. Decreasing the $\teff$ values of non solar twin stars by 20\,K gives agreement within 0.001\,mag for the mean $(V-I)_\odot$ values, but the $(V-R)_\odot$ colors still differ by 0.007\,mag. A $\teff$ decrease of about 70\,K is necessary to make the $(V-R)_\odot$ colors agree perfectly. This is highly unlikely given the high precision of our $\teff$ determinations, and therefore suggests that there are small systematic errors affecting our $(V-R)$ colors.

Even though $\logg$ is included in the regression formula (Eq.~\ref{eq:regress}) for completeness, we find that the precision of our results is not compromised if we choose to neglect it. For example, a regression using only $\teff$ and $\feh$ gives us the same solar $(B-V)$ colors of solar twins or analogs within 0.001\,mag. Moreover, the errors are identical to the case when $\logg$ is also included. This is likely the result of having selected only main-sequence stars for our sample. The impact of $\feh$ on these calculations, however, must not be ignored. A regression on $\teff$ only, or even $\teff$ and $\logg$, results in a solar $(B-V)$ color with an error that is about twice as large as that obtained using Eq.~\ref{eq:regress}. As mentioned earlier, the metallicity dependence of \ubvri\ colors is clearly seen in Figure~\ref{f:teffcolor}. We also tested regression formulae including quadratic terms, i.e., $\teff^2$, $\feh^2$, and $\logg^2$, as well as mixed terms such as $\teff\times\feh$, but found no noticeable improvements; the 1\,$\sigma$ scatter of the regression (i.e., data minus fit value residuals) did not change by more than 0.001\,mag, and the same was true for the mean values obtained for the solar colors.

\subsection{Spectral Line-Depth Ratios} \label{s:method2}

The strength of a spectral line depends on many parameters. In addition to the physical conditions of the gas in which the line is formed, which makes the line strength sensitive to the model atmosphere adopted, the properties of the atom, ion, or molecule responsible for the absorption, and those of the transition that produces the line are all directly related to the line strength. Of particular interest for our work is the excitation potential (EP) of the feature. Spectral lines with very different EP values show significantly different sensitivity to $\teff$. Thus, ratios of depths of spectral line pairs with very different EP values are known to be excellent $\teff$ indicators \cite[e.g.,][]{gray91,gray94:ldr}, and therefore they are expected to correlate well with observed colors.

\begin{deluxetable*}{cccccccc}
\tablewidth{0pc}
\tabletypesize{\footnotesize}
\tablecaption{$(B-V)_\odot$ Color Inferred from LDR Measurements}
\tablehead{\colhead{$\lambda_1$ (\AA)} & \colhead{species} & \colhead{$\lambda_2$ (\AA)} & \colhead{species} & \colhead{$N_\star$} & \colhead{$\sigma_\mathrm{fit}$} & \colhead{$(B-V)_\odot$} & \colhead{$\sigma_\mathrm{ss}$}}
\startdata
   5490.15 &  TiI & 5517.53 &  SiI & 90 & 0.014 & 0.649 & 0.005 \\
   5690.43 &  SiI & 5727.05 &   VI & 90 & 0.015 & 0.649 & 0.006 \\
   5701.11 &  SiI & 5727.05 &   VI & 90 & 0.012 & 0.651 & 0.006 \\
   5727.05 &   VI & 5753.65 &  SiI & 90 & 0.015 & 0.653 & 0.002 \\
   6007.31 &  NiI & 6046.00 &   SI & 90 & 0.013 & 0.654 & 0.005 \\
   6039.73 &   VI & 6046.00 &   SI & 90 & 0.014 & 0.645 & 0.006 \\
   6039.73 &   VI & 6052.68 &   SI & 90 & 0.013 & 0.654 & 0.004 \\
   6046.00 &   SI & 6062.89 &  FeI & 90 & 0.012 & 0.651 & 0.005 \\
   6046.00 &   SI & 6085.27 &  FeI & 42 & 0.011 & 0.652 & 0.007 \\
   6046.00 &   SI & 6091.18 &  TiI & 90 & 0.015 & 0.645 & 0.006 \\
   6052.68 &   SI & 6062.89 &  FeI & 90 & 0.012 & 0.656 & 0.003 \\
   6052.68 &   SI & 6081.44 &   VI & 34 & 0.014 & 0.654 & 0.005 \\
   6052.68 &   SI & 6091.18 &  TiI & 90 & 0.014 & 0.652 & 0.004 \\
   6090.21 &   VI & 6091.92 &  SiI & 90 & 0.013 & 0.655 & 0.004 \\
   6090.21 &   VI & 6106.60 &  SiI & 90 & 0.011 & 0.657 & 0.004 \\
   6090.21 &   VI & 6125.03 &  SiI & 90 & 0.013 & 0.654 & 0.003 \\
   6090.21 &   VI & 6131.86 &  SiI & 83 & 0.012 & 0.656 & 0.006 \\
   6091.92 &  SiI & 6128.99 &  NiI & 90 & 0.014 & 0.657 & 0.006 \\
   6106.60 &  SiI & 6119.53 &   VI & 90 & 0.011 & 0.656 & 0.004 \\
   6106.60 &  SiI & 6126.22 &  TiI & 90 & 0.013 & 0.653 & 0.003 \\
   6106.60 &  SiI & 6135.36 &   VI & 89 & 0.014 & 0.653 & 0.005 \\
   6108.12 &  NiI & 6155.14 &  SiI & 90 & 0.015 & 0.656 & 0.003 \\
   6119.53 &   VI & 6131.86 &  SiI & 83 & 0.014 & 0.651 & 0.006 \\
   6125.03 &  SiI & 6128.99 &  NiI & 90 & 0.014 & 0.656 & 0.002 \\
   6128.99 &  NiI & 6131.86 &  SiI & 83 & 0.013 & 0.658 & 0.007 \\
   6175.42 &  NiI & 6224.51 &   VI & 84 & 0.015 & 0.655 & 0.003 \\
   6176.81 &  NiI & 6224.51 &   VI & 84 & 0.015 & 0.655 & 0.003 \\
   6186.74 &  NiI & 6224.51 &   VI & 32 & 0.014 & 0.656 & 0.004 \\
   6199.19 &   VI & 6215.15 &  FeI & 74 & 0.015 & 0.656 & 0.004 \\
   6204.64 &  NiI & 6224.51 &   VI & 84 & 0.015 & 0.655 & 0.002 \\
   6204.64 &  NiI & 6243.11 &   VI & 90 & 0.013 & 0.649 & 0.003 \\
   6215.15 &  FeI & 6224.51 &   VI & 84 & 0.013 & 0.655 & 0.003 \\
   6215.15 &  FeI & 6251.82 &   VI & 90 & 0.015 & 0.655 & 0.004 \\
   6223.99 &  NiI & 6224.51 &   VI & 84 & 0.014 & 0.656 & 0.003 \\
   6223.99 &  NiI & 6243.11 &   VI & 89 & 0.014 & 0.648 & 0.006 \\
   6224.51 &   VI & 6230.09 &  NiI & 84 & 0.014 & 0.655 & 0.004 \\
   6230.09 &  NiI & 6243.11 &   VI & 90 & 0.013 & 0.651 & 0.005 \\
   6240.66 &  FeI & 6243.81 &  SiI & 90 & 0.014 & 0.654 & 0.010 \\
   6240.66 &  FeI & 6244.48 &  SiI & 90 & 0.014 & 0.654 & 0.008 \\
   6243.11 &   VI & 6243.81 &  SiI & 90 & 0.015 & 0.649 & 0.002 \\
   6327.60 &  NiI & 6414.99 &  SiI & 35 & 0.013 & 0.655 & 0.005 \\
   6414.99 &  SiI & 6498.95 &  FeI & 35 & 0.013 & 0.651 & 0.007 \\
   6710.31 &  FeI & 6748.84 &   SI & 89 & 0.012 & 0.651 & 0.006 \\
   6710.31 &  FeI & 6757.17 &   SI & 90 & 0.012 & 0.650 & 0.007 \\
   6757.17 &   SI & 6806.85 &  FeI & 34 & 0.011 & 0.656 & 0.005

\enddata
\label{t:ldr_bv}
\end{deluxetable*}

\cite{gray92:colors} was the first to use line-depth ratios (LDRs) to infer solar colors. As pointed out by him, one of the great advantages of using LDRs is that they are nearly insensitive to the stellar metallicity, at least for nearby thin-disk stars, because, to first approximation, the line strengths scale with $\feh$ regardless of the element producing the line. If, in addition, the line pairs have similar wavelengths and the spectroscopic data used are very homogeneous, particularly concerning the continuum normalization, line-depth ratios are also independent of spectral resolution. Using LDRs to infer the solar colors has also the great advantage of being a completely model-independent approach.

\begin{figure}
\epsscale{1.2}
\plotone{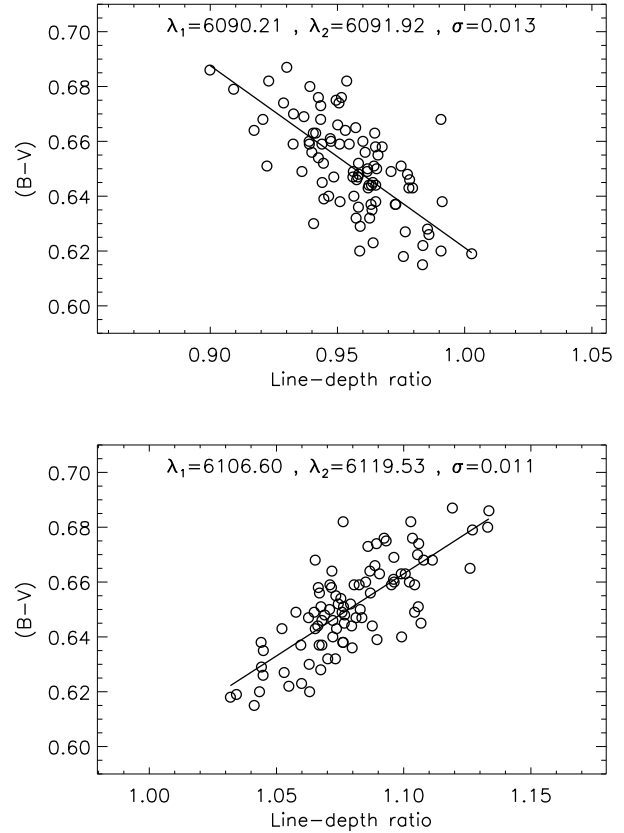}
\caption{Two examples of observed $(B-V)$ color as a function of line-depth ratio measured in our spectra. The wavelengths of the lines used and the $1\,\sigma$ of the linear fit shown with a solid line are given in the upper part of each panel.}
\label{f:ldr}
\end{figure}

We used our high resolution, high signal-to-noise ratio spectra to measure as many as possible LDRs for all line pairs listed in the study by \cite{kovtyukh03}, and inspected the LDR versus color relations obtained using our photometric data. We fitted a straight line to each of these relations, and computed the standard deviation ($1\,\sigma$) of the fit minus data residuals. Two examples of these fits are shown in Figure~\ref{f:ldr}. Then we measured the line-depth ratios in our solar spectra, which are in fact reflected Sun-light observations of bright asteroids, and used the LDR versus color fits to infer a solar color for each line pair. The weighted mean and average values obtained from all line pairs used were adopted as the final solar colors. Not all line-pairs listed in the \cite{kovtyukh03} study were used in the end. Line-pairs for which the linear fits had a $1\,\sigma$ value with a significant contribution from observational errors in the spectra, were discarded. For example, for $(B-V)$ we excluded the fits with $1\,\sigma>0.015$\,mag, because the typical $(B-V)$ error is 0.01\,mag, and adopting only the pairs with $1\,\sigma<0.015$\,mag implies that the only pairs that are used are those in which the spectroscopic errors (i.e., the errors in LDR), when propagated to the photometric data in this relation, are similar to the photometric ones, or smaller. Although somewhat arbitrary, this automated procedure eliminates line-pairs which may be affected by blends, continuum normalization issues intrinsic to our data, and/or instrumental imperfections.

As an example, in Table~\ref{t:ldr_bv} we list all the line pairs used to derive the solar $(B-V)$ color from LDR versus color relations. For each pair, we provide the number of stars, $N_\star$, used to construct the empirical relation and the standard deviation of the fit minus data residuals ($\sigma_\mathrm{fit}$). Also, for each pair we provide the mean and standard deviation of the $(B-V)$ color that corresponds to the nine reflected Sun-light asteroid observations used for solar reference ($\sigma_\mathrm{ss}$). This is because each solar spectrum gives us a slightly different value for the line-depth ratio of each pair. Note that the standard deviation from the mean color of our nine solar spectra is very small; in many cases it is below 0.005\,mag. The weighted mean and sample variance of the $(B-V)$ solar colors inferred from the 45 line pairs used is finally adopted as the solar color. We used as weights ($w$) the inverse of the standard deviations of the LDR versus color fits and the $1\,\sigma$ scatter in the colors obtained for the nine solar spectra, added in quadrature, i.e., $1/w=\sigma_\mathrm{fit}^2+\sigma_\mathrm{ss}^2$ (see Table~\ref{t:ldr_bv}).

Some of the scatter seen in Figure~\ref{f:ldr} could in principle be attributed to $\feh$ and/or $\logg$ effects. To test this hypothesis, we repeated the procedures described above, but using, instead of a simple linear fit of LDR versus color, a linear regression similar to Eq.~\ref{eq:regress}, replacing the ($\teff-5777$\,K) term with the LDR values. The exact same mean value and error was obtained for $(B-V)_\odot$, suggesting that the impact of $\feh$ and $\logg$ on the LDR versus color relations is below the 0.001\,mag level. This implies that the scatter seen in Figure~\ref{f:ldr} is dominated by observational errors in both LDR and $(B-V)$.

We computed solar colors for each line pair and for each solar spectrum available. Our spectra come from two different sources and were taken on several different observing runs. The results given in Table~\ref{t:ldr_bv} were obtained using all available data. We made sure that analyzing the data separately per run or per observing site does not improve these results in a significant manner. In fact, due to the lower number of stars available to derive the solar colors, this approach typically gives us larger errors, in some cases about twice as large for $(B-V)$, for example. Thus, we conclude that small differences in the spectral resolution, sky conditions, and/or instrumental setup have a negligible impact in our derivation of the solar colors. This observation also suggests that the continuum normalization of all our available data is robust and consistent across different data sets as well as observing runs and sites.

\begin{deluxetable}{ccc}
\tablewidth{0pc}
\tabletypesize{\small}
\tablecaption{Solar Colors Inferred from LDR Measurements}
\tablehead{\colhead{color} & \colhead{value} & \colhead{$N_\mathrm{pairs}$}}
\startdata
$(B-V)$ &$0.653\pm0.003$& 45 \\
$(U-B)$ &$0.158\pm0.009$& 42 \\
$(V-R)$ &$0.356\pm0.003$& 47 \\
$(V-I)$ &$0.701\pm0.003$& 53

\enddata
\label{t:ldr}
\end{deluxetable}

We performed a similar exercise to the one described above to derive the other \ubvri\ solar colors, which are listed in Table~\ref{t:ldr}. Using the LDR technique, we find $(B-V)_\odot=0.653\pm0.003$, $(U-B)_\odot=0.158\pm0.009$, $(V-R)_\odot=0.356\pm0.003$, and $(V-I)=0.701\pm0.003$, in excellent agreement with the solar colors obtained with the method described in Section~\ref{s:method1}. The significantly smaller error bars obtained with the LDR method are probably due to the fact that no systematic uncertainties similar to those of the stellar atmospheric parameters affect the LDR measurements, in addition to the fact that the spectroscopic data are very homogeneous and of extremely high quality.

\section{CONCLUSIONS}

The problem of the lack of important photometric data for solar analog stars in the \ubvri\ system has been addressed and solved with our \ubvri\ observations of 80 stars very similar to our Sun, for which previously obtained high resolution, high signal-to-noise ratio spectra are available. The combined use of high-quality photometric and spectroscopic data of Sun-like stars allows us to study the Sun as a star without the need to modify or design instruments specifically for the direct observation of the Sun.

We have derived the solar colors in the \ubvri\ system using two different methods. The first one uses the atmospheric parameters $\teff$, $\logg$, and $\feh$ derived using a model-dependent analysis, whereas the second method employs only measurements of spectral line-depth ratios (LDRs) and the observed photometry, thus being completely model-independent. We find excellent agreement for the solar colors derived using these two techniques. 
In particular, we derive $(B-V)_\odot=0.653\pm0.005$, but the LDR method gives a smaller error of 0.003\,mag. An uncertainty of 0.005\,mag in $(B-V)_\odot$ translates into an error of $\pm16$\,K in $\teff$ whereas a 0.003\,mag uncertainty corresponds to only 9\,K. Thus, our highly precise solar colors can be used to constrain stellar models and calibrate effective temperature scales or color-$\teff$ relations at the 10\,K level.

With respect to the recent debate in the literature concerning the solar $(B-V)$ color, our results favor the ``red'' value closer to 0.65\,mag over the ``blue'' solar color of about 0.62\,mag. Given the high quality of our photometric and spectroscopic data, as well as our careful sample selection and derivation of solar colors from the wealth of available data, we argue that our solar \ubvri\ colors are the most precise and reliable ones published to date. Along with the solar {\it uvby}-$\beta$ colors derived by \cite{melendez10}, precise and accurate solar colors in the historically most important photometric systems are now available.

\acknowledgments

I.R.'s work was performed under contract with the California Institute of Technology (Caltech) funded by NASA through the Sagan Fellowship Program. This paper uses observations made at the South African Astronomical Observatory (SAAO). J.M. acknowledges support from FAPESP (2010/17510-3), CNPq, and USP.


\setcounter{table}{4}

\clearpage
\LongTables
\tabletypesize{\tiny}
\begin{deluxetable}{lccrccccccl}
\tablewidth{0pc}
\tablecaption{Adopted Stellar Parameters and Photometry}
\tablehead{\colhead{HIP} & \colhead{$\teff$ (K)} & \colhead{$\logg$} & \colhead{$\feh$} & \colhead{$V$} & \colhead{$(B-V)$} & \colhead{$(U-B)$} & \colhead{$(V-R)$} & \colhead{$(V-I)$} & \colhead{$N_\mathrm{obs}$\tablenotemark{1}} & Source\tablenotemark{2}}
\startdata
348 & $5777\pm40$ & $4.41\pm0.07$ & $-0.130\pm0.024$ &$8.600\pm0.004$ & $0.644\pm0.008$ & $0.151\pm0.026$ & $0.348\pm0.004$ & $0.695\pm0.007$ & 3 & SPM+SAAO+OPD+LIT \\ 
996 & $5860\pm41$ & $4.38\pm0.07$ & $0.000\pm0.022$ &$8.197\pm0.012$ & $0.643\pm0.019$ & $0.146\pm0.023$ & $0.351\pm0.004$ & $0.689\pm0.006$ & 3 & SPM+SAAO+OPD \\ 
1499 & $5756\pm44$ & $4.37\pm0.05$ & $0.189\pm0.015$ &$6.474\pm0.012$ & $0.680\pm0.004$ & $0.265\pm0.011$ & $0.368\pm0.004$ & $0.714\pm0.004$ & 2 & SAAO+LIT \\ 
2131 & $5720\pm41$ & $4.38\pm0.07$ & $-0.210\pm0.026$ &$8.923\pm0.008$ & $0.643\pm0.004$ & $0.106\pm0.023$ & $0.355\pm0.009$ & $0.720\pm0.009$ & 1 & SPM+LIT \\ 
2894 & $5820\pm44$ & $4.54\pm0.07$ & $-0.030\pm0.025$ &$8.651\pm0.018$ & $0.659\pm0.028$ & $0.200\pm0.039$ & $0.350\pm0.034$ & $0.703\pm0.019$ & 1 & SPM \\ 
4909 & $5836\pm54$ & $4.44\pm0.07$ & $0.020\pm0.024$ &$8.512\pm0.006$ & $0.637\pm0.004$ & $0.119\pm0.013$ & $0.357\pm0.005$ & $0.687\pm0.004$ & 4 & SPM+SAAO+OPD \\ 
5134 & $5779\pm38$ & $4.49\pm0.07$ & $-0.190\pm0.023$ &$8.973\pm0.005$ & $0.637\pm0.007$ & $0.074\pm0.007$ & $0.346\pm0.004$ & $0.703\pm0.004$ & 4 & SPM+SAAO+OPD \\ 
6407 & $5787\pm25$ & $4.47\pm0.03$ & $-0.090\pm0.011$ &$8.624\pm0.002$ & $0.652\pm0.005$ & $0.135\pm0.010$ & $0.360\pm0.004$ & $0.704\pm0.004$ & 4 & SPM+SAAO+OPD+LIT \\ 
7245 & $5843\pm47$ & $4.53\pm0.07$ & $0.100\pm0.023$ &$8.361\pm0.001$ & $0.675\pm0.006$ & $0.148\pm0.017$ & $0.354\pm0.004$ & $0.691\pm0.004$ & 2 & SPM+OPD+LIT \\ 
8507 & $5720\pm55$ & $4.44\pm0.08$ & $-0.080\pm0.026$ &$8.899\pm0.004$ & $0.651\pm0.004$ & $0.130\pm0.007$ & $0.363\pm0.005$ & $0.730\pm0.004$ & 4 & SPM+SAAO+OPD \\ 
8841 & $5676\pm45$ & $4.50\pm0.06$ & $-0.120\pm0.021$ &$9.246\pm0.006$ & $0.674\pm0.004$ & $0.157\pm0.014$ & $0.378\pm0.004$ & $0.729\pm0.015$ & 2 & SAAO+LIT \\ 
9349 & $5825\pm28$ & $4.49\pm0.06$ & $0.010\pm0.017$ &$7.992\pm0.017$ & $0.650\pm0.004$ & $0.147\pm0.008$ & $0.343\pm0.004$ & $0.691\pm0.004$ & 3 & SPM+SAAO \\ 
11072 & $5897\pm84$ & $4.01\pm0.06$ & $-0.037\pm0.057$ &$5.190\pm0.007$ & $0.597\pm0.004$ & $0.120\pm0.004$ & $0.355\pm0.020$ & $0.692\pm0.020$ & 0 & LIT \\ 
11728 & $5738\pm30$ & $4.37\pm0.05$ & $0.045\pm0.019$ &\nodata & $0.666\pm0.015$ & \nodata & \nodata & \nodata & 0 & LIT \\ 
11915 & $5793\pm43$ & $4.45\pm0.06$ & $-0.050\pm0.021$ &$8.615\pm0.008$ & $0.649\pm0.004$ & $0.134\pm0.004$ & $0.354\pm0.004$ & $0.699\pm0.004$ & 2 & SAAO+LIT \\ 
12186 & $5812\pm34$ & $4.09\pm0.05$ & $0.094\pm0.040$ &$5.785\pm0.006$ & $0.654\pm0.007$ & $0.180\pm0.028$ & $0.360\pm0.010$ & $0.700\pm0.010$ & 0 & LIT \\ 
14614 & $5803\pm28$ & $4.47\pm0.03$ & $-0.104\pm0.016$ &$7.840\pm0.010$ & $0.620\pm0.010$ & $0.130\pm0.010$ & \nodata & \nodata & 0 & LIT \\ 
14632 & $6026\pm42$ & $4.28\pm0.05$ & $0.136\pm0.019$ &$4.047\pm0.008$ & $0.595\pm0.007$ & $0.118\pm0.010$ & \nodata & \nodata & 0 & LIT \\ 
15457 & $5771\pm65$ & $4.56\pm0.02$ & $0.078\pm0.041$ &$4.836\pm0.010$ & $0.679\pm0.007$ & $0.188\pm0.008$ & $0.384\pm0.005$ & $0.726\pm0.008$ & 0 & LIT \\ 
18261 & $5891\pm34$ & $4.44\pm0.05$ & $0.002\pm0.016$ &$7.980\pm0.027$ & $0.628\pm0.006$ & $0.083\pm0.017$ & $0.327\pm0.047$ & $0.670\pm0.028$ & 1 & SPM+LIT \\ 
22263 & $5826\pm48$ & $4.54\pm0.01$ & $0.005\pm0.029$ &$5.497\pm0.012$ & $0.632\pm0.012$ & $0.136\pm0.007$ & $0.359\pm0.005$ & $0.691\pm0.005$ & 0 & LIT \\ 
22528 & $5683\pm52$ & $4.33\pm0.10$ & $-0.350\pm0.035$ &$9.540\pm0.010$ & $0.630\pm0.010$ & $0.090\pm0.010$ & \nodata & \nodata & 0 & LIT \\ 
23835 & $5723\pm33$ & $4.16\pm0.05$ & $-0.184\pm0.017$ &$4.920\pm0.034$ & $0.645\pm0.005$ & $0.142\pm0.006$ & \nodata & \nodata & 0 & LIT \\ 
25670 & $5755\pm37$ & $4.38\pm0.05$ & $0.071\pm0.017$ &$8.275\pm0.021$ & $0.659\pm0.004$ & $0.167\pm0.012$ & $0.341\pm0.036$ & $0.698\pm0.024$ & 1 & SPM+LIT \\ 
28336 & $5713\pm61$ & $4.53\pm0.08$ & $-0.170\pm0.027$ &$8.998\pm0.001$ & $0.654\pm0.007$ & $0.108\pm0.019$ & $0.354\pm0.007$ & $0.704\pm0.010$ & 2 & SPM+SAAO+LIT \\ 
29525 & $5715\pm61$ & $4.41\pm0.04$ & $-0.005\pm0.036$ &$6.442\pm0.014$ & $0.660\pm0.015$ & $0.160\pm0.004$ & \nodata & \nodata & 0 & LIT \\ 
30037 & $5690\pm30$ & $4.42\pm0.06$ & $0.050\pm0.030$ &$9.162\pm0.015$ & $0.682\pm0.015$ & $0.213\pm0.015$ & $0.361\pm0.015$ & $0.706\pm0.015$ & 1 & SAAO \\ 
30502 & $5745\pm25$ & $4.47\pm0.05$ & $-0.010\pm0.020$ &$8.667\pm0.015$ & $0.664\pm0.015$ & $0.152\pm0.015$ & $0.368\pm0.015$ & $0.707\pm0.015$ & 1 & SAAO \\ 
36512 & $5740\pm15$ & $4.50\pm0.03$ & $-0.092\pm0.020$ &$7.729\pm0.011$ & $0.656\pm0.004$ & $0.121\pm0.004$ & $0.356\pm0.005$ & $0.696\pm0.004$ & 3 & SPM+SAAO \\ 
38072 & $5839\pm68$ & $4.53\pm0.11$ & $0.060\pm0.037$ &$9.222\pm0.002$ & $0.648\pm0.004$ & $0.147\pm0.009$ & $0.362\pm0.005$ & $0.701\pm0.004$ & 4 & SPM+SAAO \\ 
38228 & $5693\pm58$ & $4.52\pm0.07$ & $0.007\pm0.025$ &$6.900\pm0.010$ & $0.682\pm0.004$ & \nodata & \nodata & \nodata & 0 & LIT \\ 
39748 & $5835\pm30$ & $4.48\pm0.06$ & $-0.200\pm0.030$ &$8.592\pm0.009$ & $0.615\pm0.004$ & $0.050\pm0.004$ & $0.340\pm0.023$ & $0.681\pm0.004$ & 4 & SAAO+OPD \\ 
41317 & $5724\pm15$ & $4.46\pm0.03$ & $-0.044\pm0.020$ &$7.807\pm0.004$ & $0.668\pm0.004$ & $0.158\pm0.004$ & $0.365\pm0.005$ & $0.712\pm0.008$ & 3 & SPM+SAAO+LIT \\ 
42438 & $5864\pm47$ & $4.46\pm0.09$ & $-0.052\pm0.026$ &$5.631\pm0.009$ & $0.619\pm0.004$ & $0.070\pm0.004$ & \nodata & \nodata & 0 & LIT \\ 
43190 & $5775\pm30$ & $4.37\pm0.06$ & $0.120\pm0.030$ &$8.508\pm0.015$ & $0.670\pm0.015$ & $0.232\pm0.015$ & $0.370\pm0.015$ & $0.696\pm0.015$ & 1 & SAAO \\ 
44324 & $5934\pm49$ & $4.51\pm0.06$ & $-0.020\pm0.019$ &$7.943\pm0.010$ & $0.620\pm0.027$ & $0.083\pm0.035$ & $0.321\pm0.011$ & $0.674\pm0.016$ & 4 & SPM \\ 
44713 & $5784\pm35$ & $4.36\pm0.03$ & $0.096\pm0.024$ &$7.306\pm0.006$ & $0.668\pm0.005$ & $0.201\pm0.006$ & $0.371\pm0.005$ & $0.713\pm0.005$ & 0 & LIT \\ 
44935 & $5800\pm25$ & $4.41\pm0.05$ & $0.070\pm0.020$ &$8.739\pm0.014$ & $0.645\pm0.004$ & $0.200\pm0.007$ & $0.344\pm0.004$ & $0.690\pm0.004$ & 2 & SPM+SAAO \\ 
44997 & $5782\pm29$ & $4.52\pm0.04$ & $0.033\pm0.020$ &$8.347\pm0.023$ & $0.659\pm0.008$ & $0.202\pm0.005$ & $0.348\pm0.008$ & $0.698\pm0.014$ & 2 & SPM+SAAO \\ 
46066 & $5709\pm65$ & $4.49\pm0.12$ & $-0.070\pm0.039$ &$8.928\pm0.007$ & $0.664\pm0.017$ & $0.188\pm0.016$ & $0.360\pm0.008$ & $0.717\pm0.008$ & 2 & SPM \\ 
46126 & $5890\pm30$ & $4.48\pm0.06$ & $0.140\pm0.030$ &$8.514\pm0.006$ & $0.651\pm0.004$ & $0.149\pm0.030$ & $0.354\pm0.006$ & $0.704\pm0.023$ & 2 & SAAO+LIT \\ 
49572 & $5831\pm52$ & $4.33\pm0.06$ & $0.010\pm0.021$ &$9.288\pm0.006$ & $0.640\pm0.007$ & $0.138\pm0.006$ & $0.336\pm0.008$ & $0.702\pm0.007$ & 1 & SPM \\ 
49756 & $5804\pm52$ & $4.45\pm0.07$ & $0.041\pm0.023$ &$7.540\pm0.008$ & $0.647\pm0.004$ & $0.185\pm0.004$ & $0.343\pm0.004$ & $0.687\pm0.007$ & 4 & SPM+SAAO+OPD+LIT \\ 
51258 & $5720\pm25$ & $4.23\pm0.05$ & $0.360\pm0.030$ &$7.874\pm0.004$ & $0.725\pm0.004$ & $0.344\pm0.008$ & $0.386\pm0.006$ & $0.735\pm0.004$ & 2 & SAAO+LIT \\ 
52137 & $5842\pm69$ & $4.56\pm0.08$ & $0.070\pm0.026$ &$8.640\pm0.010$ & $0.640\pm0.010$ & $0.190\pm0.010$ & \nodata & \nodata & 0 & LIT \\ 
53721 & $5916\pm53$ & $4.48\pm0.01$ & $0.027\pm0.038$ &$5.049\pm0.015$ & $0.606\pm0.010$ & $0.124\pm0.007$ & \nodata & \nodata & 0 & LIT \\ 
54102 & $5870\pm30$ & $4.51\pm0.06$ & $0.040\pm0.030$ &$8.653\pm0.004$ & $0.649\pm0.004$ & $0.142\pm0.015$ & $0.346\pm0.004$ & $0.698\pm0.004$ & 2 & SAAO \\ 
55409 & $5760\pm25$ & $4.52\pm0.05$ & $-0.010\pm0.020$ &$8.002\pm0.014$ & $0.659\pm0.004$ & $0.193\pm0.011$ & $0.368\pm0.004$ & $0.719\pm0.004$ & 5 & SAAO+OPD+LIT \\ 
55459 & $5838\pm21$ & $4.42\pm0.03$ & $0.038\pm0.012$ &$7.646\pm0.004$ & $0.644\pm0.004$ & $0.147\pm0.010$ & $0.338\pm0.006$ & $0.692\pm0.006$ & 1 & SPM+LIT \\ 
56948 & $5795\pm23$ & $4.43\pm0.03$ & $0.023\pm0.014$ &$8.669\pm0.004$ & $0.646\pm0.006$ & $0.180\pm0.006$ & $0.339\pm0.006$ & $0.680\pm0.006$ & 1 & SPM \\ 
56997 & $5559\pm65$ & $4.53\pm0.08$ & $-0.030\pm0.027$ &$5.321\pm0.015$ & $0.723\pm0.013$ & $0.261\pm0.018$ & \nodata & \nodata & 0 & LIT \\ 
57291 & $5690\pm22$ & $4.30\pm0.04$ & $0.304\pm0.030$ &$7.466\pm0.008$ & $0.740\pm0.004$ & $0.354\pm0.006$ & $0.375\pm0.016$ & $0.732\pm0.013$ & 2 & SAAO \\ 
59357 & $5810\pm30$ & $4.45\pm0.06$ & $-0.240\pm0.030$ &$8.731\pm0.039$ & $0.618\pm0.004$ & $0.078\pm0.004$ & $0.351\pm0.010$ & $0.715\pm0.031$ & 6 & SPM+SAAO+OPD+LIT \\ 
59610 & $5899\pm62$ & $4.34\pm0.04$ & $-0.034\pm0.041$ &$7.360\pm0.010$ & $0.640\pm0.010$ & \nodata & \nodata & \nodata & 0 & LIT \\ 
60081 & $5811\pm21$ & $4.38\pm0.04$ & $0.315\pm0.030$ &$8.023\pm0.007$ & $0.696\pm0.006$ & $0.290\pm0.008$ & $0.373\pm0.007$ & $0.702\pm0.007$ & 2 & SAAO \\ 
60314 & $5874\pm72$ & $4.52\pm0.10$ & $0.110\pm0.033$ &$8.780\pm0.008$ & $0.649\pm0.017$ & $0.155\pm0.007$ & $0.337\pm0.011$ & $0.676\pm0.009$ & 1 & SPM+LIT \\ 
60370 & $5897\pm25$ & $4.46\pm0.05$ & $0.171\pm0.030$ &$6.703\pm0.004$ & $0.641\pm0.005$ & $0.148\pm0.013$ & $0.349\pm0.008$ & $0.674\pm0.008$ & 2 & SAAO+LIT \\ 
60653 & $5725\pm30$ & $4.38\pm0.06$ & $-0.290\pm0.030$ &$8.735\pm0.005$ & $0.638\pm0.014$ & $0.109\pm0.015$ & $0.360\pm0.004$ & $0.702\pm0.010$ & 4 & SAAO+OPD \\ 
62175 & $5849\pm51$ & $4.43\pm0.06$ & $0.140\pm0.021$ &$8.011\pm0.005$ & $0.661\pm0.004$ & $0.194\pm0.007$ & $0.345\pm0.006$ & $0.682\pm0.006$ & 1 & SPM+LIT \\ 
64150 & $5755\pm41$ & $4.39\pm0.05$ & $0.056\pm0.016$ &$6.822\pm0.061$ & $0.676\pm0.020$ & $0.204\pm0.004$ & $0.374\pm0.013$ & $0.700\pm0.012$ & 3 & SPM+SAAO+LIT \\ 
64497 & $5860\pm110$ & $4.56\pm0.11$ & $0.120\pm0.037$ &$8.978\pm0.057$ & $0.668\pm0.022$ & $0.182\pm0.008$ & $0.362\pm0.009$ & $0.687\pm0.004$ & 6 & SPM+SAAO+OPD \\ 
64713 & $5815\pm25$ & $4.52\pm0.05$ & $-0.010\pm0.020$ &$9.260\pm0.022$ & $0.649\pm0.004$ & $0.140\pm0.010$ & $0.351\pm0.023$ & $0.710\pm0.019$ & 5 & SPM+SAAO+OPD \\ 
64794 & $5743\pm61$ & $4.33\pm0.08$ & $-0.100\pm0.027$ &$8.428\pm0.041$ & $0.637\pm0.006$ & $0.141\pm0.008$ & $0.344\pm0.028$ & $0.701\pm0.013$ & 6 & SPM+SAAO+OPD+LIT \\ 
64993 & $5875\pm30$ & $4.56\pm0.06$ & $0.090\pm0.030$ &$8.900\pm0.022$ & $0.656\pm0.008$ & $0.166\pm0.008$ & $0.348\pm0.013$ & $0.706\pm0.009$ & 4 & SPM+SAAO+OPD \\ 
66618 & $5951\pm25$ & $4.35\pm0.05$ & $0.135\pm0.030$ &$6.962\pm0.004$ & $0.622\pm0.004$ & $0.175\pm0.005$ & \nodata & \nodata & 0 & LIT \\ 
66885 & $5685\pm30$ & $4.48\pm0.06$ & $-0.380\pm0.030$ &$9.302\pm0.014$ & $0.635\pm0.004$ & $0.077\pm0.014$ & $0.364\pm0.014$ & $0.730\pm0.005$ & 4 & SPM+SAAO+OPD+LIT \\ 
69063 & $5670\pm30$ & $4.31\pm0.06$ & $-0.450\pm0.030$ &$8.882\pm0.006$ & $0.623\pm0.004$ & $0.068\pm0.004$ & $0.352\pm0.004$ & $0.706\pm0.010$ & 2 & SAAO+LIT \\ 
71683 & $5840\pm22$ & $4.33\pm0.04$ & $0.228\pm0.030$ &$0.002\pm0.008$ & $0.653\pm0.023$ & $0.230\pm0.004$ & $0.362\pm0.010$ & $0.693\pm0.010$ & 0 & LIT \\ 
72659 & $5517\pm67$ & $4.56\pm0.09$ & $-0.117\pm0.033$ &$4.718\pm0.008$ & $0.748\pm0.019$ & $0.231\pm0.019$ & \nodata & \nodata & 0 & LIT \\ 
73815 & $5803\pm33$ & $4.34\pm0.05$ & $0.020\pm0.016$ &$8.174\pm0.003$ & $0.663\pm0.006$ & $0.164\pm0.005$ & $0.352\pm0.009$ & $0.696\pm0.006$ & 5 & SPM+SAAO+OPD+LIT \\ 
74341 & $5853\pm57$ & $4.51\pm0.08$ & $0.090\pm0.026$ &$8.860\pm0.013$ & $0.673\pm0.013$ & $0.165\pm0.018$ & $0.348\pm0.010$ & $0.689\pm0.007$ & 6 & SPM+OPD \\ 
74389 & $5859\pm24$ & $4.48\pm0.04$ & $0.105\pm0.030$ &$7.773\pm0.014$ & $0.636\pm0.014$ & $0.153\pm0.014$ & $0.349\pm0.004$ & $0.687\pm0.005$ & 6 & SPM+SAAO+OPD \\ 
75923 & $5775\pm25$ & $4.56\pm0.05$ & $-0.020\pm0.020$ &$9.156\pm0.012$ & $0.658\pm0.006$ & $0.137\pm0.004$ & $0.353\pm0.013$ & $0.704\pm0.015$ & 6 & SPM+SAAO+OPD \\ 
77052 & $5697\pm33$ & $4.54\pm0.02$ & $0.035\pm0.023$ &$5.868\pm0.011$ & $0.686\pm0.011$ & $0.234\pm0.010$ & $0.380\pm0.010$ & $0.740\pm0.010$ & 0 & LIT \\ 
77466 & $5700\pm56$ & $4.40\pm0.09$ & $-0.280\pm0.028$ &$9.204\pm0.009$ & $0.647\pm0.005$ & $0.120\pm0.014$ & \nodata & \nodata & 0 & LIT \\ 
77740 & $5900\pm19$ & $4.45\pm0.04$ & $0.125\pm0.030$ &\nodata & $0.628\pm0.012$ & \nodata & \nodata & \nodata & 0 & LIT \\ 
77883 & $5695\pm25$ & $4.39\pm0.05$ & $0.040\pm0.020$ &$8.755\pm0.020$ & $0.687\pm0.008$ & $0.224\pm0.005$ & $0.371\pm0.004$ & $0.729\pm0.014$ & 6 & SPM+SAAO+OPD \\ 
78028 & $5879\pm98$ & $4.57\pm0.12$ & $-0.030\pm0.041$ &$8.651\pm0.012$ & $0.638\pm0.019$ & $0.118\pm0.024$ & $0.334\pm0.016$ & $0.683\pm0.016$ & 5 & SPM \\ 
78680 & $5923\pm67$ & $4.57\pm0.08$ & $-0.000\pm0.027$ &$8.243\pm0.013$ & $0.626\pm0.018$ & $0.079\pm0.021$ & $0.337\pm0.016$ & $0.698\pm0.016$ & 3 & SPM \\ 
79186 & $5709\pm48$ & $4.27\pm0.08$ & $-0.120\pm0.024$ &$8.341\pm0.014$ & $0.676\pm0.004$ & $0.140\pm0.033$ & $0.356\pm0.022$ & $0.724\pm0.018$ & 3 & SPM+LIT \\ 
79304 & $5945\pm30$ & $4.53\pm0.06$ & $0.110\pm0.030$ &$8.703\pm0.021$ & $0.646\pm0.015$ & $0.160\pm0.009$ & $0.347\pm0.004$ & $0.683\pm0.008$ & 7 & SPM+SAAO+OPD \\ 
79578 & $5860\pm33$ & $4.53\pm0.07$ & $0.072\pm0.030$ &$6.533\pm0.033$ & $0.647\pm0.007$ & $0.145\pm0.012$ & $0.352\pm0.008$ & $0.699\pm0.004$ & 2 & SAAO+LIT \\ 
79672 & $5822\pm9$ & $4.45\pm0.02$ & $0.051\pm0.020$ &$5.510\pm0.009$ & $0.650\pm0.004$ & $0.177\pm0.004$ & $0.357\pm0.005$ & $0.691\pm0.011$ & 3 & SPM+SAAO+LIT \\ 
80337 & $5881\pm33$ & $4.53\pm0.02$ & $0.033\pm0.022$ &$5.391\pm0.012$ & $0.628\pm0.011$ & $0.108\pm0.048$ & $0.353\pm0.005$ & $0.681\pm0.006$ & 0 & LIT \\ 
81512 & $5790\pm58$ & $4.46\pm0.07$ & $-0.020\pm0.025$ &$9.245\pm0.015$ & $0.652\pm0.017$ & $0.140\pm0.019$ & $0.353\pm0.019$ & $0.712\pm0.016$ & 3 & SPM \\ 
82853 & $5640\pm30$ & $4.21\pm0.06$ & $-0.180\pm0.030$ &$8.965\pm0.163$ & $0.660\pm0.007$ & $0.181\pm0.004$ & $0.396\pm0.004$ & $0.729\pm0.008$ & 3 & SAAO+OPD \\ 
83601 & $6071\pm43$ & $4.38\pm0.08$ & $0.048\pm0.028$ &$6.013\pm0.008$ & $0.575\pm0.005$ & $0.041\pm0.012$ & $0.325\pm0.010$ & $0.635\pm0.010$ & 0 & LIT \\ 
83707 & $5880\pm30$ & $4.45\pm0.06$ & $0.150\pm0.030$ &$8.605\pm0.011$ & $0.655\pm0.004$ & $0.181\pm0.004$ & $0.350\pm0.004$ & $0.699\pm0.004$ & 3 & SAAO+OPD \\ 
85042 & $5692\pm37$ & $4.39\pm0.02$ & $0.037\pm0.026$ &$6.287\pm0.004$ & $0.679\pm0.004$ & $0.233\pm0.008$ & $0.364\pm0.020$ & $0.707\pm0.051$ & 2 & SAAO+LIT \\ 
85272 & $5700\pm30$ & $4.42\pm0.06$ & $-0.340\pm0.030$ &$9.121\pm0.000$ & $0.632\pm0.016$ & $0.095\pm0.012$ & $0.363\pm0.009$ & $0.717\pm0.006$ & 3 & SAAO+OPD \\ 
85285 & $5730\pm30$ & $4.43\pm0.06$ & $-0.390\pm0.030$ &$8.365\pm0.015$ & $0.629\pm0.011$ & $0.065\pm0.013$ & $0.348\pm0.009$ & $0.712\pm0.008$ & 7 & SPM+SAAO+OPD \\ 
86796 & $5809\pm22$ & $4.28\pm0.04$ & $0.298\pm0.030$ &$5.124\pm0.006$ & $0.700\pm0.004$ & $0.240\pm0.004$ & $0.385\pm0.004$ & $0.708\pm0.004$ & 2 & SAAO+LIT \\ 
88194 & $5735\pm21$ & $4.40\pm0.03$ & $-0.071\pm0.010$ &$7.101\pm0.035$ & $0.639\pm0.008$ & $0.126\pm0.004$ & $0.360\pm0.006$ & $0.712\pm0.008$ & 5 & SPM+OPD+LIT \\ 
88427 & $5810\pm57$ & $4.42\pm0.07$ & $-0.160\pm0.025$ &$9.329\pm0.006$ & $0.638\pm0.013$ & $0.089\pm0.022$ & $0.336\pm0.018$ & $0.704\pm0.007$ & 1 & SPM \\ 
89162 & $5835\pm30$ & $4.32\pm0.06$ & $0.070\pm0.030$ &$8.902\pm0.005$ & $0.658\pm0.005$ & $0.176\pm0.004$ & $0.360\pm0.006$ & $0.696\pm0.006$ & 4 & SAAO+OPD \\ 
89443 & $5796\pm73$ & $4.48\pm0.12$ & $-0.020\pm0.038$ &$8.843\pm0.004$ & $0.660\pm0.007$ & $0.147\pm0.013$ & $0.359\pm0.006$ & $0.715\pm0.006$ & 1 & SPM \\ 
89650 & $5855\pm25$ & $4.48\pm0.05$ & $0.020\pm0.020$ &$8.944\pm0.001$ & $0.643\pm0.004$ & $0.126\pm0.004$ & $0.356\pm0.004$ & $0.688\pm0.009$ & 7 & SAAO+OPD \\ 
91332 & $5775\pm25$ & $4.20\pm0.05$ & $0.206\pm0.030$ &$7.971\pm0.008$ & $0.692\pm0.007$ & $0.263\pm0.004$ & $0.365\pm0.015$ & $0.705\pm0.004$ & 2 & SAAO+LIT \\ 
96402 & $5713\pm49$ & $4.33\pm0.03$ & $-0.029\pm0.030$ &$7.560\pm0.010$ & $0.678\pm0.007$ & $0.154\pm0.010$ & \nodata & \nodata & 0 & LIT \\ 
96895 & $5808\pm39$ & $4.33\pm0.05$ & $0.097\pm0.020$ &$5.959\pm0.009$ & $0.644\pm0.006$ & $0.189\pm0.009$ & \nodata & \nodata & 0 & LIT \\ 
96901 & $5737\pm28$ & $4.34\pm0.04$ & $0.055\pm0.016$ &$6.228\pm0.019$ & $0.663\pm0.005$ & $0.191\pm0.016$ & \nodata & \nodata & 0 & LIT \\ 
100963 & $5802\pm17$ & $4.45\pm0.03$ & $0.008\pm0.013$ &$7.089\pm0.021$ & $0.651\pm0.013$ & $0.128\pm0.024$ & $0.342\pm0.004$ & $0.703\pm0.007$ & 3 & SPM+OPD \\ 
100970 & $5823\pm40$ & $4.23\pm0.03$ & $0.083\pm0.025$ &$6.895\pm0.015$ & $0.645\pm0.005$ & $0.180\pm0.020$ & \nodata & \nodata & 0 & LIT \\ 
109110 & $5817\pm60$ & $4.46\pm0.03$ & $0.062\pm0.030$ &$7.570\pm0.010$ & $0.674\pm0.015$ & \nodata & \nodata & \nodata & 0 & LIT \\ 
102152 & $5737\pm47$ & $4.35\pm0.06$ & $-0.010\pm0.022$ &$9.208\pm0.015$ & $0.669\pm0.004$ & $0.176\pm0.019$ & $0.382\pm0.004$ & $0.726\pm0.004$ & 3 & SPM+SAAO \\ 
104504 & $5836\pm48$ & $4.50\pm0.06$ & $-0.160\pm0.022$ &$8.544\pm0.021$ & $0.622\pm0.004$ & $0.068\pm0.012$ & $0.363\pm0.004$ & $0.701\pm0.004$ & 5 & SPM+SAAO+OPD+LIT \\ 
107350 & $6015\pm50$ & $4.48\pm0.07$ & $-0.020\pm0.019$ &$5.942\pm0.011$ & $0.587\pm0.004$ & $0.032\pm0.004$ & \nodata & \nodata & 0 & LIT \\ 
108708 & $5875\pm51$ & $4.51\pm0.07$ & $0.150\pm0.024$ &$8.939\pm0.003$ & $0.659\pm0.004$ & $0.162\pm0.012$ & $0.368\pm0.004$ & $0.711\pm0.004$ & 4 & SPM+OPD \\ 
108996 & $5838\pm56$ & $4.50\pm0.08$ & $0.060\pm0.027$ &$8.881\pm0.012$ & $0.648\pm0.008$ & $0.152\pm0.025$ & $0.351\pm0.011$ & $0.691\pm0.011$ & 6 & SPM+SAAO+OPD \\ 
109931 & $5739\pm74$ & $4.29\pm0.08$ & $0.040\pm0.026$ &$8.956\pm0.019$ & $0.663\pm0.006$ & $0.194\pm0.016$ & $0.367\pm0.034$ & $0.710\pm0.020$ & 1 & SPM+LIT \\ 
113357 & $5803\pm47$ & $4.38\pm0.05$ & $0.221\pm0.017$ &$5.467\pm0.020$ & $0.665\pm0.012$ & $0.233\pm0.028$ & \nodata & \nodata & 0 & LIT \\ 
118159 & $5905\pm44$ & $4.55\pm0.07$ & $-0.010\pm0.022$ &$9.010\pm0.006$ & $0.627\pm0.004$ & $0.090\pm0.004$ & $0.341\pm0.007$ & $0.680\pm0.004$ & 4 & SPM+SAAO+OPD

\enddata
\tablenotetext{1}{Number of photometric observations made in this work.}
\tablenotetext{2}{If the source includes LIT, which corresponds to previously published values, the LIT flag applies only to the UBV data. The RI$_\mathrm{(C)}$ data are, to the best of our knowledge, published here for the first time, except for stars with the following HIP numbers: 1499, 11072, 12186, 15457, 22263, 41317, 44713, 71683, 77052, 79672, 80337, 83601, and 86796.}
\label{t:adopted}
\end{deluxetable}
\clearpage

\end{document}